\documentclass[%
 aip,
 amsmath,amssymb,
 reprint,%
]{revtex4-1}

\usepackage{graphicx}
\usepackage[colorlinks=true, linkcolor=blue, citecolor=blue, urlcolor=blue]{hyperref}
\usepackage{hyperref}
\usepackage{dcolumn}
\usepackage{bm}
\usepackage{color}
\usepackage[utf8]{inputenc}
\usepackage[T1]{fontenc}
\usepackage{mathptmx}
\usepackage{etoolbox}
\usepackage{comment}
\usepackage[english]{babel}
\usepackage{threeparttable}
\usepackage{tabularx}
\usepackage[left=1.5cm, right=1.5cm, top=1.785cm, bottom=2.0cm]{geometry}
\newcommand{\lt}{<}

\makeatletter
\def\@email#1#2{%
 \endgroup
 \patchcmd{\titleblock@produce}
  {\frontmatter@RRAPformat}
  {\frontmatter@RRAPformat{\produce@RRAP{*#1\href{mailto:#2}{#2}}}\frontmatter@RRAPformat}
  {}{}
}%
\makeatother
\begin{document}

\preprint{AIP/123-QED}

\title{Automatic Forward Model Parameterization with Bayesian Inference of Conformational Populations}
\author{Robert M. Raddi}\noaffiliation
\author{Tim Marshall}\noaffiliation
\author{Vincent A. Voelz\textsuperscript{*}}\noaffiliation
\affiliation{
Department of Chemistry, Temple University, Philadelphia, PA 19122, USA. \\
*Corresponding author: vvoelz@temple.edu.
}%

\date{\today}

\begin{abstract}
To quantify how well theoretical predictions of structural ensembles agree with experimental measurements, we depend on the accuracy of forward models. These models are computational frameworks that generate observable quantities from molecular configurations based on empirical relationships linking specific molecular properties to experimental measurements. Bayesian Inference of Conformational Populations (BICePs) is a reweighting algorithm that reconciles simulated ensembles with ensemble-averaged experimental observations, even when such observations are sparse and/or noisy. This is achieved by sampling the posterior distribution of conformational populations under experimental restraints as well as sampling the posterior distribution of uncertainties due to random and systematic error.  In this study, we enhance the algorithm for the refinement of empirical forward model (FM) parameters.  We introduce and evaluate two novel methods for optimizing FM parameters. The first method treats FM parameters as nuisance parameters, integrating over them in the full posterior distribution. The second method employs variational minimization of a quantity called the BICePs score that reports the free energy of ``turning on'' the experimental restraints. This technique, coupled with improved likelihood functions for handling experimental outliers, facilitates force field validation and optimization, as illustrated in recent studies (Raddi et al. 2023, 2024). Using this approach, we refine parameters that modulate the Karplus relation, crucial for accurate predictions of $J$-coupling constants based on dihedral angles ($\phi$) between interacting nuclei. We validate this approach first with a toy model system, and then for human ubiquitin, predicting six sets of Karplus parameters for ${^{3}\!J}_{H^{N} H^{\alpha}}$, ${^{3}\!J}_{H^{\alpha} C^{\prime}}$, ${^{3}\!J}_{H^{N} C^{\beta}}$, ${^{3}\!J}_{H^{N} C^{\prime}}$, ${^{3}\!J}_{C^{\prime}C^{\beta}}$, ${^{3}\!J}_{C^{\prime}C^{\prime}}$.   Finally, we demonstrate that our framework naturally generalizes optimization to any differentiable forward model, such as those constructed with neural networks. This approach provides a promising direction for training and validating neural network-based forward models.
\end{abstract}

\maketitle

\section{Introduction}
In the field of molecular modeling and dynamics, the accuracy of theoretical predictions is crucial.  The quantitative agreement between theory and experiment is typically characterized through the use of accurate \textit{forward models}—computational frameworks that predict observable quantities from molecular configurations. These models often depend on empirical relationships that link specific molecular properties to experimental measurements.

Model validation and refinement of structural ensembles against NMR observables critically depends on reliable forward models (FMs) that have been robustly parameterized to reduce error. In refining these FMs, it is important to consider random and systematic errors inherent to the experimental data, which in turn contribute to overall uncertainty of integrative models.  It is usually challenging to obtain complete knowledge of such errors; instead, such errors typically need to be inferred from the data.

A further challenge is presented by missing or insufficient examples of known structures that can be used to train forward models. For NMR observables that depend on backbone $\phi$-angles, such as $J$-coupling constants, the reference data from X-ray crystallography may be missing or dynamically averaged, creating large uncertainties in the correct $\phi$-angles.  Numerous approaches have been developed to address some of these challenges. \cite{frohlking2023simultaneous, wang1996determination, habeck2005bayesian, schmidt1999self}  Some algorithms rely heavily on X-ray crystal structure data; others have many hyperparameters that need to be determined.

To address these challenges, we extend the Bayesian Inference of Conformational Populations (BICePs) algorithm\cite{voelz2021reconciling,raddi2025model} to refine FM parameters.  BICePs is a reweighting algorithm to refine structural ensembles against sparse and/or noisy experimental observables, which has been used in many previous applications.\cite{Voelz:2014fga,wan2020reconciling,hurley2021,raddi2023biceps}  BICePs infers all possible sources of error by sampling the posterior distribution of these parameters directly from the data through MCMC sampling. BICePs also computes a free energy-like quantity called the BICePs score that can be used for model selection and model parameterization.\cite{ge2018model,raddi2025model,raddi2024automated}

Recently, BICePs was enhanced with a replica-averaging forward model, making it a maximum-entropy (MaxEnt) reweighting method, and unique in that no adjustable regularization parameters are required to balance experimental information with the prior.\cite{raddi2025model}  With this new approach, the BICePs score becomes a powerful objective function to parameterize optimal models.  Here, we show that the BICePs score, which reflects the total evidence for a model, can be used for variational optimization of FM parameters.  The BICePs score contains a form of inherent regularization, and is particularly powerful when used with specialized likelihood functions that allow for the automatic detection and down-weighting of the importance of experimental observables subject to systematic error.\cite{raddi2025model}

Beyond structural biology, Bayesian methods are widely used in survival analysis to model censored and heterogeneous discrete data. Recent work by Akhtar and colleagues shows that flexible priors and robust loss functions enable accurate parameter estimation and outlier resistance in such settings \cite{akhtar2024analyzing,akhtar2025geometric}. These developments highlight the versatility of Bayesian inference for complex, noisy datasets and provide additional examples of analogous strategies for handling heterogeneity and outliers in our domain.

To effectively refine FM parameters, we use BICePs to sample over the entire posterior distribution of FM parameters, enabling simultaneous ensemble reweighting and parameter refinement. Additionally, we show that by variational minimization of the BICePs score, we obtain the same result and show that the two approaches are equivalent, with each method requiring particular considerations. We first demonstrate our method’s effectiveness using a toy model and then apply it to optimize six distinct sets of Karplus parameters for the human protein ubiquitin. As a proof-of-concept for more complex models, we also train a neural network to learn $J$-coupling constants from dihedral angles using the BICePs score as a loss function. We then discuss how this idea could be applied to neural networks for chemical shift prediction.\cite{han2024accurate,yang2021predicting,meiler2003proshift,shen2010sparta}  Of course, our framework naturally extends to other empirical forward models with tunable parameters—for example, hydrogen-deuterium exchange (HDX) models with adjustable hydrogen-bonding and burial weights, or paramagentic relaxation enhancement (PRE) models with system-specific correlation times and scaling factors. These examples illustrate the generality and flexibility of our approach for refining a wide range of forward models to improve the agreement between simulations and experiment.

\section{Theory}

In this section, we develop two different theoretical approaches to optimize forward models. In the first method, we include forward model parameters in the posterior, and sample the full posterior distribution of all parameters.  In the second approach, we optimize the forward model by minimizing the BICePs score, a free energy-like quantity that can be used for variational optimization.  While the two methods are theoretically equivalent (see the supplemental material for a short proof of this), each has practical advantages and disadvantages. 

\subsection{Forward model optimization by posterior sampling of forward model parameters}

BICePs uses a Bayesian statistical framework, inspired by Inferential Structure Determination (ISD) \cite{rieping2005inferential}, to model the posterior distribution $p(X,\sigma)$, for conformational states $X$, and nuisance parameters $\sigma$, which characterize the extent of uncertainty in the experimental observables $D$:
\begin{equation}
  p(X,\sigma | D) \propto  p(D | X,\sigma) p(X) p(\sigma).
  \label{eq:early_biceps}
\end{equation}
Here, $p(D | X,\sigma)$ is a likelihood function that uses a forward model to enforce the experimental restraints, $p(X)$ is a prior distribution of conformational populations from some theoretical model, and $p(\sigma) \sim \sigma^{-1}$ is a non-informative Jeffrey's prior.

For a specific forward model $g(X, \theta)$ with a set of FM parameters $\theta = (\theta_1, \theta_2,...,\theta_m)$, where $m$ is the number of parameters, we can additionally include the parameters in the posterior,
\begin{equation}
  p(X,\sigma, \theta | D) \propto  p(D | X,\sigma,\theta) p(X) p(\sigma) p(\theta).
  \label{eq:posterior}
\end{equation}

\paragraph*{Replica-averaging.}
When BICePs is used with a replica-averaged forward model, it becomes a MaxEnt reweighting method in the limit of large numbers of replicas \cite{pitera2012use,cavalli2013molecular,cesari2018using,roux2013statistical,hummer2015bayesian,bonomi2016metainference}. Consider a set of $N$ replicas, $\mathbf{X} = \{X_{r}\}$, where $X_r$ is the conformational state being sampled by replica $r$.  To compare the sampled replicas with ensemble-averaged experimental observables, we define a replica-averaged forward model $g(\mathbf{X},\theta) = \frac{1}{N}\sum^{N}_{r} g(X_{r},\theta)$. This quantity is an estimator of the true ensemble average, with an error due to finite sampling for the observable $j$ estimated using the standard error of the mean (SEM):\cite{pitera2012use,bonomi2016metainference} $\sigma^{\text{SEM}}_{j} = \sqrt{\frac{1}{N}\sum\nolimits^{N}_{r} (g_{j}(X_{r},\theta) - \langle g_{j}(\mathbf{X},\theta) \rangle )^{2}}$.  Thus, $\sigma^{\text{SEM}}_{j}$ decreases as the square root of the number of replicas.

For a single forward model with parameters $\theta$, the  joint posterior distribution for all parameters is
\begin{equation}
  p(X, \sigma, \theta | D) \propto  \prod_{r=1}^{N} p(X_{r})   p(D | g(\mathbf{X}, \theta), \sigma)  p(\sigma) p(\theta)
  \label{eq:joint_posterior}
\end{equation}
where $\mathbf{X}$ is a set of $N$ conformation replicas.   In many cases, we wish to optimize several forward models at once.  For example, there are unique sets of Karplus parameters in the forward models for vicinal $J$-coupling constant observables ${^{3}\!J}_{H^{N} H^{\alpha}}$, ${^{3}\!J}_{H^{\alpha} C^{\prime}}$, ${^{3}\!J}_{H^{N} C^{\beta}}$,  etc.   If there are $K$ different forward models $\theta^{(1)},\theta^{(2)},  ... \theta^{(K)}$, each with their own set of parameters and error distributions,  we can describe their full joint posterior distribution as
\begin{equation}
\begin{split}
& p(X, \sigma, \theta^{(1)},\theta^{(2)},  ... \theta^{(K)} | D) \\
 &\propto  \prod_{r=1}^{N} p(X_{r})  \prod_{k=1}^{K} p(D_{k} | g(\mathbf{X}, \theta^{(k)}), \sigma_{k})  p(\sigma_{k}) p(\theta^{(k)})
  \end{split}
\end{equation}
where $D_k$ are the experimental observables predicted by the $k^{th}$ forward model,  $\theta^{(k)}$ are the parameters for the $k^{th}$ FM, and   $\sigma_{k}= \sqrt{ (\sigma^{\text{SEM}}_{k})^{2} + (\sigma^{\text{B}}_{k})^{2}}$ describing the total error for the $k^{th}$ observable. The total error originates from both finite sampling through the standard error of the mean $(\sigma^{\text{SEM}}_{k})$, and uncertainty in the experimental measurements, known as a Bayesian uncertainty parameter $\sigma^{\text{B}}_{k}$. The prior distribution of uncertainties $p(\sigma_{k})$ is treated as a non-informative Jeffrey's prior $(\sigma_{k}^{-1})$ for each collection of observables, and the posterior of FM parameters $p(\theta | D)$ is recovered by marginalization over all $X$ and $\sigma$:
\begin{equation}
  p(\theta | D) = \sum_{X} \int p(X, \sigma, \theta | D) d \sigma
\end{equation}

\paragraph*{Gradients speed up convergence.}
In our methodology, Markov chain Monte Carlo (MCMC) is used to sample the posterior with acceptances following the Metropolis-Hastings criterion. Our algorithm can be used with or without gradients. However, significantly faster convergence, especially in higher dimensions, is achieved through an integration of stochastic gradient descent approach. Our gradient descent approach allows for informed updates to the FM parameters, incorporating stochastic noise to facilitate the escape from local minima and enhance exploration of the parameter space.

The update mechanism is succinctly encapsulated in the equation:
\begin{equation}
  \theta_{\text{trial}} = \theta_{\text{old}} - l_{\text{rate}} \cdot \nabla u + \eta \cdot \mathcal{N}(0, 1)
\end{equation}
where $\theta_{\text{trial}}$ and $\theta_{\text{old}}$ denote the trial parameters and previous parameters, respectively. The learning rate is denoted by $l_{\text{rate}}$, $\nabla u$ signifies the calculated gradient of BICePs energy function $u = -\ln p(X, \sigma, \theta | D)$ with respect to the parameters $\theta$, and $\eta$ scales the noise drawn from a standard normal distribution $\mathcal{N}(0, 1)$.

This parameter update strategy is designed to satisfy the Metropolis-Hastings criterion, ensuring that each step in the parameter space not only moves towards minimizing the energy of the forward model but also adheres to the probabilistic acceptance of potentially non-optimal moves to avoid local optima traps. Ergodic sampling is ensured by "turning off" the gradient after burn-in. The sampling procedure involves: (1) acquiring derivatives of the FM parameters, (2) perturbing these parameters based on the derived information, (3) predicting observables using perturbed FM parameters and computing the total energy, and (4) assessing the new energy against the previous to determine acceptance based on the Metropolis-Hastings criterion. This ensures a thorough and effective search of the parameter space, taking advantage of both the landscape topology and stochastic elements to guide the exploration.

\paragraph*{The Good-and-Bad model accounts for systematic error due to outlier measurements.}
Recent work has developed sophisticated likelihood models for BICePs, designed to detect outliers, that are robust in the presence of systematic error\cite{raddi2025model}. Raddi et al. (2025) developed two likelihood models for this purpose: the Student's model and the Good-and-Bad model.  In related work using BICePs to perform automated force field optimization, we used the Student's model.\cite{raddi2024automated}  Here, in this work, we use the Good-and-Bad model, primarily because the derivatives are less complicated than the Student's model.

The Good-and-Bad likelihood model\cite{raddi2025model} assumes that the level of noise is mostly uniform, except for a few erratic measurements. This limits the number of uncertainty parameters that need to be sampled, while still capturing outliers.  Consider a model in which the uncertainties $\sigma_j$ for particular observables $j$ are distributed with respect to some typical uncertainty $\sigma^{B}$ according to a conditional probability $p(\sigma_j | \sigma^{B})$. We derive a posterior for the parameters of the $k^{th}$ FM having a single uncertainty parameter $\sigma^{B}$ by marginalizing over all $\sigma_j$,
\begin{widetext}
 \begin{equation}
p(\mathbf{X},\sigma_{0}, \theta^{(k)}|D)  \propto \prod^{N}_{r=1} p(X_{r}) \prod^{N_{d}}_{j=1} \int\limits_{\sigma^{SEM}}^{\infty} p(d_{j}|g_{j}(\mathbf{X}, \theta^{(k)}), \sigma_{j}) p(\sigma_{j} | \sigma_{0}) d\sigma_{j}
 \label{eq:posterior_marginal_over_sigma}
\end{equation}
 where  $\sigma_{0} = \sqrt{(\sigma^{B})^{2} + (\sigma^{SEM})^{2}}$. Under the Good-Bad model, we say that the "good" data consist of observables normally distributed about their true values with effective variance $\sigma_{0}^{2}$, while the "bad" data are subject to systematic error, leading to a larger effective variance $\varphi^{2}\sigma_{0}^{2}$, where $\varphi \geq 1$.

  By this assignment,  $p(\sigma_{j} | \sigma_{0})$ from equation \ref{eq:posterior_marginal_over_sigma} becomes
\begin{equation}
p(\sigma_{j} | \sigma_{0}, \omega, \varphi) = \omega \delta(\sigma_{j} - \varphi \sigma_{0}) + (1 - \omega)\delta(\sigma_{j} - \sigma_{0})
\end{equation}
where  $0 \leq \omega < 1$ describes the fraction of "bad" observables.  Since the value of $\omega$ is unknown, it is treated as a nuisance parameter, and marginalized over its range. The resulting posterior is
\begin{equation}
\begin{split}
 &p(\mathbf{X}, \sigma_{0}, \varphi, \theta^{(k)} | D) \\
  &\propto  \prod^{N}_{r=1} \left\{ p(X_{r})  \prod^{N_{d}}_{j=1} \int\limits_{0}^{1} d\omega \int\limits_{\sigma^{\text{SEM}}}^{\infty}  \exp\left( - \frac{(d_{j} - g_{j}(\mathbf{X}, \theta^{(k)}))^{2} }{2\sigma_{j}^{2}} \right)
    \frac{\omega \delta(\sigma_{j} - \varphi \sigma_{0}) + (1 - \omega)\delta(\sigma_{j} - \sigma_{0})}{\sqrt{2\pi}\sigma_{j}} d\sigma_{j} \right\} \\
&=  \prod^{N}_{r=1} \left\{ p(X_{r}) \prod^{N_{d}}_{j=1} \Biggl( \frac{1-H(\sigma^{SEM} - \sigma_{0})}{2}\mathcal{N}\left( g_{j}(\mathbf{X},\theta^{(k)}) \,\middle|\, d_j, \sigma_{0}^2 \right) + \frac{1-H(\sigma^{SEM} - \varphi\sigma_{0})}{2\varphi} \mathcal{N}\left(g_{j}(\mathbf{X},\theta^{(k)}) \,\middle|\, d_j, \varphi^2 \sigma_{0}^2 \right)  \Biggr) \right\},
\end{split}
  \label{eq:GB_post}
\end{equation}
 where $H$ is the Heaviside step function.  After marginalization, we are left with the Bayesian uncertainty parameter $\sigma_{0}^{B}$, and an additional parameter $\varphi$. Both parameters are sampled in the posterior. When $\varphi=1$, the model reverts to a Gaussian likelihood model. When considering the full posterior, this extra nuisance parameter is given a non-informative Jeffrey's prior,  $p(\varphi) \sim \varphi^{-1}$.

For a single set of FM parameters (for simplicity), the BICePs energy function, $u = -\log p(\mathbf{X}, \sigma_{0}, \varphi, \theta^{(k)} | D)$, the negative logarithm of the posterior in its full form is given by
\begin{equation}
\begin{split}
  \hspace{-0.25cm}
u = \sum^{N}_{r=1} -\log(p(X_{r})) - N\sum^{N_{d}}_{j=1} \log\left[ \frac{1-H(\sigma^{SEM} - \sigma_{0})}{2}\mathcal{N}\left(g_{j}(\mathbf{X},\theta^{(k)}) \,\middle|\, d_j, \sigma_{0}^2 \right) + \frac{1-H(\sigma^{SEM} - \varphi\sigma_{0})}{2\varphi} \mathcal{N}\left(g_{j}(\mathbf{X},\theta^{(k)}) \,\middle|\, d_j, \varphi^2 \sigma_{0}^2 \right) \right].
\end{split}
  \label{eq:energy_GB}
\end{equation}
When $\varphi=1$,  this energy function is the same the result we would get if we used a Gaussian likelihood function,
\begin{equation}
  \begin{split}
    u = \sum^{N}_{r=1} -\log(p(X_{r})) + N  \left[ \sum_{j=1}^{N_{d}} - \log \left( \frac{1}{\sqrt{2 \pi} \sigma_{0}} \right) + \frac{(d_{j}-g_{j}(\mathbf{X} , \theta))^{2}}{2 \sigma_{0}^{2}} - \log ( p(\sigma_{0}) ) \right].
  \label{eq:energy_Gaussian}
\end{split}
\end{equation}

The first derivative of equation \ref{eq:energy_GB} with respect to the $i^{th}$ FM parameter in the parameter set $\theta^{(k)}$ is
\begin{equation}
  \frac{\partial u}{\partial \theta^{(k)}_{i}} = N \sum_{j=1}^{N_d} \frac{\partial g_{j}(\mathbf{X},\theta^{(k)})}{\partial\theta^{(k)}_{i} } \, \frac{(d_{j} - g_{j}(\mathbf{X},\theta^{(k)})) }{ \sigma_{0}^{2}  } \, \left( \frac{\varphi^{2} \left(1 - H\left(\sigma^{SEM} - \sigma_{0}\right)\right) \mathcal{N}_{\varphi^{2}}^{-1} + \left(1 - H\left(- \varphi \sigma_{0} + \sigma^{SEM}\right)\right) \mathcal{N}_{1}^{-1}}{ \varphi^2 \left(H\left(\sigma^{SEM} - \sigma_{0}\right) - 1\right) \mathcal{N}_{\varphi^{2}}^{-1} + \varphi^{2}\left(H\left(- \varphi \sigma_{0} + \sigma^{SEM}\right) - 1\right) \mathcal{N}_{1}^{-1} } \right),
  \label{eq:GB_derivative}
\end{equation}
where $\mathcal{N}_{1}^{-1} = \mathcal{N}\left(g_{j}(\mathbf{X},\theta^{(k)}) \,\middle|\, d_j, \sigma_{0}^2 \right)^{-1}$ and $\mathcal{N}_{\varphi^{2}}^{-1} = \mathcal{N}\left(g_{j}(\mathbf{X},\theta^{(k)}) \,\middle|\, d_j, \varphi^2 \sigma_{0}^2 \right)^{-1}$. Both represent the reciprocal of Gaussian likelihoods, where $\mathcal{N}_{1}^{-1}$ and $\mathcal{N}_{\varphi^{2}}^{-1}$ denote inverse-weighted forms of normal distributions with variances $\sigma_0^2$ and $\varphi^2 \sigma_0^2$, respectively.
\end{widetext}
When $\varphi=1$, the gradient becomes
\begin{equation}
  \frac{\partial u}{\partial \theta^{(k)}_{i}} = - N  \sum_{j=1}^{N_{d}} \frac{\partial g_{j}(\mathbf{X},\theta^{(k)})}{\partial \theta^{(k)}_{i}}\frac{(d_{j}-g_{j}(\mathbf{X},\theta^{(k)}))}{ \sigma_0^2} .
  \label{eq:Gaussian_derivative}
\end{equation}

Second derivatives of the BICePs energy function and the BICePs score are useful for descent and uncertainty quantification using other forward models.  We refrain from writing out the second derivative here, but the derivation is straightforward.  For the forward models we consider below (Karplus relations), all have second derivatives that go to zero. For more general cases, see Appendix \ref{sec:SI_theory} for more details. The energy of the Good-Bad likelihood model and its first and second derivatives are shown in Figure \ref{fig:GB_derivatives}.

\subsection{Forward model optimization by variational minimization of the BICePs score}

As we show in Results and Discussion, treating the forward model parameters as nuisance parameters and sampling them as part of the full posterior is an efficient strategy to refine FM parameters.  However, in the limit of a very large number of FM parameters, the dimensionality of the posterior may become unwieldy and may suffer from the ``curse of dimensionality''.

As an alternative strategy, we consider optimization of forward model parameters using variational minimization of the BICePs score, an approach that we have used to perform automated force field optimization\cite{raddi2024automated}. In this approach, the FM parameters are no longer part of the joint posterior density. Instead, the posterior is conditioned on the set of FM parameters $\theta$, that is, equation \ref{eq:posterior} becomes
\begin{equation}
  p(X,\sigma| D, \theta ) \propto  p(D, \theta  | X,\sigma) p(X) p(\sigma)
\end{equation}
In this view, ensemble refinement is performed with a static set of FM parameters for each BICePs calculation.  The BICePs score then enables the selection of the optimal model parameterized by $\theta$.

BICePs evaluates model quality by calculating a free energy-like quantity called the BICePs score.  For a forward model with parameters $\theta$, the BICePs score $f(\theta)$ is computed as the negative logarithm of a Bayes factor comparing the total evidence of a given model against a well-defined reference, marginalizing over all uncertainty,
\begin{equation}
  f(\theta) = - \ln \left(Z(\theta) \big/ Z_{0}\right) ,
  \label{eq:biceps_score}
\end{equation}
where
\begin{equation}
  Z(\theta) = \iint \exp\left(-u(\mathbf{X}, \mathbf{\sigma}; \theta)\right) d \mathbf{X} d \mathbf{\sigma}
  \label{eq:model_evidence}
\end{equation}
is the evidence for FM parameters $\theta$, $Z_{0}$ is the evidence for a suitable reference state, and $u$ is the unchanged BICePs energy function (equation \ref{eq:energy_GB}).  To construct the reference state, we consider a series of likelihoods $p_{\xi}(D, \theta|\mathbf{X},\mathbf{\sigma}) \sim [p(D|\mathbf{X},\mathbf{\sigma})]^{\xi}$ parameterized by $\xi \in [0,1]$, and set the reference state as the thermodynamic ensemble corresponding to $\xi=0$.  The BICePs score is then calculated as the free energy of "turning on" experimental restraints ($\xi=0\rightarrow 1$).

It should be noted that in other applications of BICePs,\cite{raddi2025model,raddi2024automated} the reference state for the BICePs score is defined using the $\lambda=0$ state for a series of priors $p_{\lambda}(X) \sim [p(X)]^{\lambda}$, and the BICePs score is computed as the free energy of ($\lambda=0\rightarrow 1$) and ($\xi=0\rightarrow 1$) transformations.  Here, since we are only interested in evaluating and/or parameterizing the likelihood functions, we set $p(X)$ to be uniform.

To optimize forward model parameters, the BICePs score is used as an objective function that is estimated given a candidate value of $\theta$. Based on the estimate (and possibly other previous estimates),  new candidates for $\theta$ are iteratively proposed and accepted or rejected, with the goal of finding the values $\theta^{*}$ that minimize the BICePs score.

For this purpose, gradient-based searches are very helpful. Conveniently, the derivative of the BICePs score $f(\theta)$ with respect to the FM parameters reduces to the Boltzmann-averaged values of the gradient $\partial u / \partial \theta$ :
\begin{equation}
\begin{split}
  \frac{\partial f(\theta)}{\partial \theta_{i}} &= \iint \frac{1}{Z(\theta)}  \left[ \frac{\partial u}{\partial \theta_{i}} \right] \exp \left( - u \right)  d \mathbf{X} d \mathbf{\sigma} = \bigg\langle \frac{\partial u}{\partial \theta_{i}} \bigg\rangle
\end{split}
  \label{eq:biceps_score_gradient}
\end{equation}
In our results below, we use the first-order optimization method  L-BFGS-B,\cite{zhu1997} which only requires first derivatives. For more complex forward models, second-order methods may be needed.  The Hessian matrix (of second partial derivatives of the BICePs score with respect to FM parameters) can be used for uncertainty estimation when $f(\theta)$ is convex.  Assuming this is true, we estimate uncertainties in the FM model parameters using covariances obtained through inversion of the Hessian. Conceptually, this means that the estimated uncertainties in the best-fit values represent the widths of the basins in the BICePs score landscape. For non-convex problems, this can sometimes lead to an under-estimation of uncertainty depending upon the curvature of the basins of the local minima in the BICePs score landscape.

Calculation of the BICePs score (a free energy difference) and its derivatives (expectation values of energy derivatives) is performed using the MBAR free energy estimator,\cite{Shirts:2008eza} by sampling at several intermediates  $\xi= 0\rightarrow 1$, which enables accurate estimates of all quantities. The accuracy of the BICePs score depends on converged sampling and sufficient thermodynamic overlap of intermediates ($\xi= 0\rightarrow 1$) in the BICePs computation. To ensure strong overlap, we optimize the $\xi$-values by spacing ensembles equidistantly in thermodynamic length,\cite{sivak2012thermodynamic,shenfeld2009minimizing} employing a strategy akin to the "thermodynamic trailblazing" method proposed by Rizzi et al.\cite{rizzi2020improving} This optimization is performed using an algorithm we have developed called \texttt{pylambdaopt}.\cite{novack2025simple}



\section{Results}

\subsection*{A toy model to test the performance of forward model optimization}\label{sec:toy_model}
\begin{figure}[!htb]
  \includegraphics[width=\linewidth]{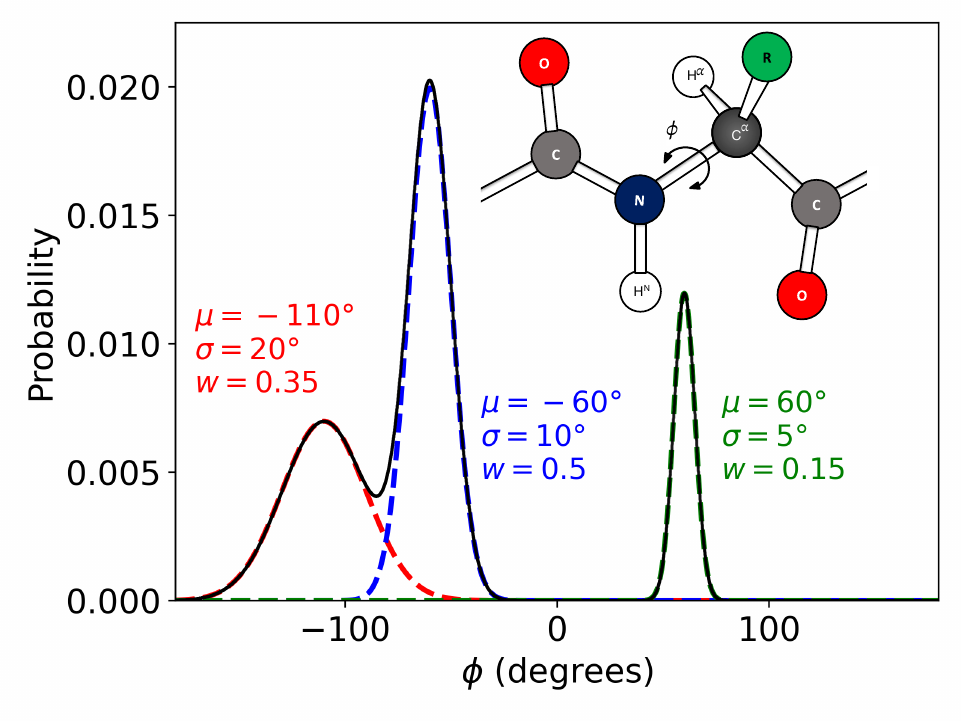}
  \caption{\small \textbf{A toy model for measuring the performance of forward model optimization.}
  The $\phi$-angles for each conformational state is pulled from a multi-modal distribution and corresponding energies. (a) This multi-modal distribution of $\phi$-angles was intended to represent configurations with different secondary structure elements having three distinct modes described by the mean ($\mu$), standard deviation ($\sigma$) and weight ($w$):  beta sheets ($\mu=-110^{\circ}$, $\sigma=20^{\circ}$, $w=0.35$), right-handed helices ($\mu=-60^{\circ}$, $\sigma=10^{\circ}$, $w=0.5$), and left-handed helices ($\mu=60^{\circ}$, $\sigma=5^{\circ}$, $w=0.15$). (b) Cartoon representation of the backbone torsion angle, $\phi$.
  }
  \label{fig:toy_model}
\end{figure}

To investigate the performance of BICePs in optimizing forward models, we introduce a simplified, yet physical, toy model. This model is designed to mimic the complexity of protein backbone structures by generating $\phi$-angles from a multi-modal distribution, thereby emulating configurations characteristic of different secondary structural elements (Figure \ref{fig:toy_model}).  This distribution encompasses three distinct modes, each characterized by a mean ($\mu$), standard deviation ($\sigma$), and weight ($w$): beta sheets ($\mu=-110^{\circ}$, $\sigma=20^{\circ}$, $w=0.35$), right-handed helices ($\mu=-60^{\circ}$, $\sigma=10^{\circ}$, $w=0.5$), and left-handed helices ($\mu=60^{\circ}$, $\sigma=5^{\circ}$, $w=0.15$). These parameters were chosen to accurately reflect the structural variability found in proteins.  Angles $\phi_i$ were sampled from the multi-modal distribution,
\begin{equation}
p(\boldsymbol{\phi}| \mathbf{\mu}, \mathbf{\sigma}) = \sum_{l} w_{l} \frac{1}{\sqrt{2\pi{\sigma_{l}}^{2}}} \exp \left( - \frac{(\phi - \mu_{l})^{2}}{2{\sigma_{l}}^{2}} \right).
\end{equation}
The sampled $\phi_i$ were then used to calculate synthetic ensemble-averaged experimental $J$-coupling constants $^{3}J(\phi)$, using the Karplus relation
\begin{equation}
  ^{3}\!J(\phi) = A \cos^{2}(\phi) + B\cos (\phi) + C,
\end{equation}
with the coefficients set to their chosen $\textit{true}$ values:  ($A^{*}$=6.51, $B^{*}$=-1.76, $C^{*}$=1.6).  The synthetic $J$-coupling constant observables were calculated using the ensemble-average, as $\sum_{i}{^{3}J(\phi_i}) p(\phi_i)$.

\subsection*{Including forward model parameters in the BICePs posterior is a robust method to optimize Karplus parameters in the presence of experimental errors.}\label{sec:toy_model}

Using the toy model described above, we next perform tests to examine whether including forward model parameters in the BICePs posterior can be used to find the optimal Karplus parameters starting from an initial reference forward model with parameters $(A_{0},B_{0},C_{0})$.  Through these tests, we evaluate the precision of the results, and how robust the algorithm is in the presence of noise.

To evaluate the robustness of our algorithm in the presence of experimental inaccuracies, we introduced random and systematic errors of varying magnitudes ($\sigma_{\text{data}}$) into the synthetic experimental scalar couplings. The performance of our Good-Bad likelihood model, a Gaussian likelihood model, and singular value decomposition (SVD) was compared under these conditions.

\paragraph*{SVD calculations.}  Parameters of many reference Karplus relations were previously determined using a singular value determination approach.\cite{Hu1997DeterminationO} To test this approach against our BICePs method, we derived Karplus parameters $\theta=\{$A$, $B$,$C$\}$ using a weighted singular value decomposition (SVD) fitting approach to optimally fit the $J$-coupling values as a function of dihedral angles. For each observation $j$ across $N_{d}$ measurements, the matrix $M$ was constructed with rows for each $\phi$ angle:
\begin{equation}
  M = \begin{bmatrix}
  \sum_{X} p(X)\cos^2(\phi_{1,X} + \phi_{0}) & \sum_{X} p(X)\cos(\phi_{1,X} + \phi_{0}) & 1 \\
  \sum_{X} p(X)\cos^2(\phi_{2,X} + \phi_{0}) & \sum_{X} p(X)\cos(\phi_{2,X} + \phi_{0}) & 1 \\
  \vdots & \vdots & \vdots \\
  \sum_{X} p(X)\cos^2(\phi_{N_{d},X} + \phi_{0}) & \sum_{X} p(X)\cos(\phi_{N_{d},X} + \phi_{0}) & 1 \\
  \end{bmatrix}
\end{equation}
where $p(X)$ represents the true populations for state $X$, and $\phi_0$ is the phase shift of $-60^\circ$.

SVD was applied to decompose the matrix as $M=U \Sigma V^{T}$, and Karplus coefficients were derived using:
\begin{equation}
\theta = V (\Sigma + \epsilon I)^{-1} U^T J^\text{exp},
\end{equation}
where $\epsilon=1 \times 10^{-6}$ is a small regularization term added to the diagonal of $\Sigma$ to ensure stability of the pseudo-inverse, and $J^\text{exp}$ represents the vector of experimental $J$-coupling values. This method ensures robust estimation of $\theta$ under ideal experimental conditions, given the true conformational populations. In practice, the true populations are not known \textit{a priori}. The uncertainty in SVD coefficients was determined through 1000 iterations of fitting, each omitting 10\% of the data points chosen at random.

Typical uncertainties in NMR frequency measurements range from 0.1 to 1.0 Hz, primarily influenced by magnetic field strength, instrument quality, sample conditions, and the specifics of the pulse sequence used.
In these experiments, 100 conformational states and 60 synthetic experimental scalar couplings were used. We introduced systematic error by shifting the experimental ${^{3}\!J}$ values by +2.0 Hz to +4.0 Hz for up to 20\% of the data points. BICePs calculations were performed by averaging FM parameters over three chains of MCMC starting from different initial parameters ($\{A=9, B=-1, C=1\},\{A=4, B=0, C=3\}, \{A=0, B=0, C=0\}$). Regardless of different starting parameters, posterior sampling universally converges to "true" optimal FM parameters.  In these calculations, we used 32 BICePs replicas, and burned 10k steps followed by 50k steps of MCMC sampling.

We evaluated model performance by the root-mean-square error (RMSE) between the true $J$-coupling values with parameters $\{A^{*}=6.51, B^{*}=-1.76, C^{*}=1.6\}$ and the $J$-coupling values using predicted Karplus coefficients for all 60 synthetic measurements, performed over 1k independent trials of random generations of toy model data. Average RMSE results, computed over 100 BICePs calculations, highlight the algorithm's robustness and its ability to accurately predict FM parameters even in the presence of data perturbations. Error bars in our results represent the standard deviation across these calculations, providing a comprehensive measure of the algorithm's reliability under various experimental accuracy.

\begin{figure}[!ht]
  \includegraphics[width=\linewidth]{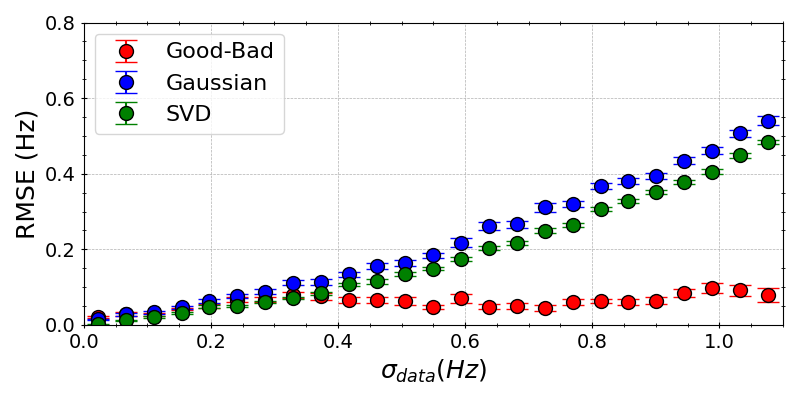}
  \caption{\small Comparative analysis in the performance of the Good-Bad likelihood model (red), a Gaussian likelihood model (blue), and singular value decomposition (SVD) using the "true" $\phi$ angles with synthetic experimental data. Here, we induced random and systematic error of varying magnitude ($\sigma_{\text{data}}$) to the experimental scalar couplings.  Model performance was measured by computing RMSE (Hz) between the "true" scalar couplings and the couplings generated from the Karplus relations with predicted Karplus coefficients over 1,500 random perturbations to the experimental data, and represent the average of 100 BICePs calculations. Error bars represent the standard deviation.  Predictions from SVD and the Gaussian likelihood model become notably less dependable when data incorporates errors, especially when $\sigma_{\text{data}}$ exceeds 0.5 Hz.}
  \label{fig:rand_and_sys_error}
\end{figure}

Our findings indicate that the Good-Bad likelihood model (red) exhibits superior insensitivity to experimental errors compared to a traditional Gaussian likelihood model (blue) and SVD (green) approaches (Figure \ref{fig:rand_and_sys_error}). Predictions from SVD and the Gaussian likelihood model become notably less dependable when the experimental data contains errors, especially when $\sigma_{\text{data}}$ exceeds 0.5 Hz. On average, error in predictions (RMSE) from the Good-Bad model does not exceed 0.1 Hz over the full range of $\sigma_{\text{data}}$.

An example of a single trial of forward model parameter refinement using the toy model is shown in Figure \ref{fig:toy_example}, where BICePs predicts Karplus coefficients by posterior sampling over FM parameters. Both BICePs and SVD methods successfully reproduce the "true" Karplus curve. However, BICePs can accurately identify the error present in the data ($\sigma_{\text{data}} = 0.471$ Hz), as indicated in the marginal posterior of uncertainty $p(\sigma_{J})$. The BICePs-predicted maximum a posteriori uncertainty was found to be $\sigma_{J} = 0.272$ with a variance scaling parameter of $\varphi_{J} = 1.98$. The marginal posterior distributions of FM parameters for the Good-Bad model were $\{A=6.6 \pm 0.04, B=-1.8 \pm 0.02, C=1.5 \pm 0.03\}$, and for SVD, $\{A=6.11 \pm 0.06, B=-1.63 \pm 0.04, C=1.80 \pm 0.04\}$.

In addition to the Good-bad model, we refined parameters using the Student's model ($\{A=6.8 \pm 0.03, B=-1.9 \pm 0.03, C=1.4 \pm 0.03 \}$) to demonstrate that the Student's model yields similar performance (Figure \ref{fig:students_toy_example}). The computed Gelman-Rubin ($\hat{R}$) statistic\cite{gelman1992inference} for these calculations was found to be $\hat{R} = 1.01$ for each of the marginal posterior distributions of Karplus coefficients, which demonstrates that our chains converge to the same parameter location with similar variance.

Furthermore, we assessed model performance across varying qualities of prior structural ensembles as illustrated in Figure \ref{fig:toy_biceps_score_corr}.
By introducing varying levels of prior error $\sigma_{\text{prior}}$ (measured in degrees) through perturbations to the "true" $\phi$ angles, even in the presence of random and systematic error, we observed strong correlation between the BICePs score and the quality of the structural ensemble, with a coefficient of determination $R^{2}$ of $0.99$. For these calculations, we employed the Good-Bad model, utilizing 32 replicas, and conducted 1,000 random perturbations to the $\phi$ angles with errors up to $\sigma_{\text{prior}} = 4^\circ$, and perturbations to the experimental data $\sigma_{\text{data}} = 0.68 \pm 0.24$ Hz.


The comprehensive evaluation of our algorithm with this toy model underscores its efficacy in accurately determining FM parameters, reflecting scenarios commonly encountered in real-world applications. The robust performance of the algorithm, even in the face of random and systematic errors, can be attributed to BICePs' sophisticated error-handling within its likelihood models. This approach also ensures that predicted FM parameters derived from sub-optimal structural ensembles remain reliable. Additionally, our findings reveal a strong correlation between the BICePs score and the quality of the structural ensemble, demonstrating an immense utility in this context.

\subsection*{Variational minimization of the BICePs score to optimize FM parameters and comparison with posterior sampling}

Next, we performed variational minimization of the BICePs score to determine optimal FM parameters for our toy model (Figure \ref{fig:toy_example}) , and compared the results to the posterior sampling approach under the same conditions.

Variational minimization was performed using the Good-Bad model with 4 replicas (for reduced computational cost), where each evaluation of the objective function consisted of running 10,000 MCMC steps.  Optimal parameters were determined to be $\{A=6.31 \pm 0.02, B=-1.69 \pm 0.03, C=1.69 \pm 0.01 \}$, averaged over 3 independent runs with very low variance between runs, shown in Figure \ref{fig:variational_BS_toy_model}. Regardless of different starting parameters ($\{A=9, B=-1, C=1\},\{A=4, B=0, C=3\}, \{A=0, B=0, C=0\}$), variational minimization converges to ``true'' optimal FM parameters.  This analysis demonstrated that both the joint posterior sampling approach and variational minimization show nearly equivalent performance when applied to this model.

As a method for forward model optimization, variational minimization of the BICePs score has advantages and disadvantages. This method is particularly advantageous for handling many FM parameters, offering a potential solution to the curse of dimensionality faced by basic Monte Carlo Markov Chain (MCMC) methods. Additionally, it is easier for users to adapt different forward models, and performs exceptionally well in convex landscapes. When landscapes are non-convex, however, the inverse Hessian may not provide a comprehensive view of the parameter space's uncertainty; instead, uncertainty estimation could be computed using the variance across multiple BICePs runs starting from different initial parameters. The variational minimization approach also requires careful consideration of disperse starting parameters to ensure global minimization.

The posterior sampling method, which involves sampling the joint posterior distribution of forward model (FM) parameters, has several advantages. One significant benefit is that the posterior distribution provides a direct estimate of the uncertainties of the parameters and their covariance. Compared to variational minimization, this method generally has a faster runtime and is particularly effective in handling non-convex landscapes, allowing for robust parameter estimation even in complex scenarios. However, it is not without drawbacks. As the number of FM parameters increases, the posterior sampling method may encounter the curse of dimensionality, which makes it computationally challenging to explore the parameter space efficiently.

In summary, while both approaches are valuable tools for parameter estimation in parameter and ensemble refinement, each has its strengths and weaknesses. The choice between these methods should be guided by the specific characteristics of the problem at hand, such as the landscape's convexity and the number of parameters involved.

\subsection*{Determination of optimal Karplus coefficients for ubiquitin}
As an application of our algorithm, we applied BICePs to human ubiquitin to predict Karplus coefficients for six sets of scalar coupling constants: ${^{3}\!J}_{H^{N} H^{\alpha}}$, ${^{3}\!J}_{H^{\alpha} C^{\prime}}$, ${^{3}\!J}_{H^{N} C^{\beta}}$, ${^{3}\!J}_{H^{N} C^{\prime}}$, ${^{3}\!J}_{C^{\prime}C^{\beta}}$, and ${^{3}\!J}_{C^{\prime}C^{\prime}}$. To test the robustness of our algorithm, we conducted a comprehensive evaluation using three different structural ensembles as priors, each derived from distinct computational approaches: (1) 10 conformations from the NMR-refined structural ensemble, 1D3Z\cite{cornilescu1998validation}, (2) 144 conformations from NMR-restrained simulations, 2NR2\cite{richter2007mumo}, and (3) 25 conformations from the RosettaFold2 (RF2) algorithm.\cite{baek2023efficient} Further details about these ubiquitin structural ensembles are given in Supporting Information.

We validated the forward model parameters derived from each prior using the BICePs score, $R^2$ and mean absolute errors (MAE) for forward model predictions.  As priors for these validation calculations, we used three independent structural ensembles: 1D3Z and 2NR2 (described above), and a 500-state conformational ensemble derived from a millisecond-long simulation of ubiquitin using CHARMM22*.\cite{Piana2013} Further details about the CHARMM22* ensemble is given in Supporting Information.

To refine the forward model (FM) parameters, we employed full joint posterior distribution sampling. This method was chosen to navigate the non-convex parameter space efficiently, given its relatively low dimensionality (18 FM parameters). Estimates of FM parameters were made by averaging the BICePs results over four Markov Chain Monte Carlo (MCMC) sampling runs, each starting from distinct initial parameters: $\{A=9, B=-1, C=1\}$, $\{A=4, B=0, C=3\}$, $\{A=0, B=0, C=0\}$, and $\{A=6, B=-1, C=0\}$. Flexible residues were excluded from the calculations, consistent with previous studies\cite{wang1996determination,habeck2005bayesian}. As a result, a total of 346 $J$-couplings were used in these refinements. We used the Good-Bad model with 32 BICePs replicas, discarding the first 50k steps as burn-in, followed by 50k steps for MCMC sampling. Unlike the parameters derived from 1D3Z and RF2, the Karplus coefficients obtained by using the 2NR2 ensemble required a burn-in of 100k steps to appropriately converge due to a larger number of conformational states. The six sets of refined Karplus coefficients resulting from the 1D3Z, 2NR2 and RF2 ensembles are presented in Table \ref{tab:karplus_parameters}.
\begin{table}[!htb]
\begin{threeparttable}
  \caption{Coefficients for the Karplus relation $^{3}\!J(\phi) = A \cos^{2}(\phi + \phi_{0}) + B \cos (\phi + \phi_{0}) + C$, determined by BICePs sampling the joint posterior of FM parameters.}
\label{tab:karplus_parameters}
\centering
\small
\begin{tabularx}{\linewidth}{lcrrrr}
\hline
  & & $\phi_{0}$ & $\mathrm{A}$ (Hz) & $\mathrm{B}$ (Hz) & $\mathrm{C}$ (Hz) \\
\hline
  {\large ${^{3}\!J}_{C^{\prime}C}$          } & 1 & $0^{\circ}$    & $1.71 \pm 0.02$ & $-0.85 \pm 0.01$ & $0.54 \pm 0.00$  \\
                                               & 2 & $0^{\circ}$    & $1.30 \pm 0.03$ & $-0.91 \pm 0.01$ & $0.62 \pm 0.01$  \\
                                               & 3 & $0^{\circ}$    & $1.62 \pm 0.03$ & $-0.87 \pm 0.01$ &  $0.63 \pm 0.01$  \\
  {\large ${^{3}\!J}_{C^{\prime}C^{\beta}}$  } & 1 & $60^{\circ}$   & $1.83 \pm 0.04$ & $0.34 \pm 0.05$  & $0.41 \pm 0.02$  \\
                                               & 2 & $60^{\circ}$   & $2.20 \pm 0.04$ & $0.34 \pm 0.04$  & $0.04 \pm 0.02$  \\
                                               & 3 & $60^{\circ}$   & $1.81 \pm 0.04$ & $0.38 \pm 0.04$  &  $0.31 \pm 0.02$  \\
  {\large ${^{3}\!J}_{H^{\alpha} C^{\prime}}$} & 1 & $120^{\circ}$  & $3.64 \pm 0.02$ & $-2.14 \pm 0.02$ & $1.27 \pm 0.02$  \\
                                               & 2 & $120^{\circ}$  & $4.10 \pm 0.03$ & $-2.00 \pm 0.02$ & $0.95 \pm 0.02$  \\
                                               & 3 & $120^{\circ}$  & $3.78 \pm 0.02$ & $-2.12 \pm 0.02$ &  $1.21 \pm 0.02$  \\
  {\large ${^{3}\!J}_{H^{N} C^{\prime}}$     } & 1 & $180^{\circ}$  & $4.33 \pm 0.04$ & $-1.17 \pm 0.01$ & $0.14 \pm 0.01$  \\
                                               & 2 & $180^{\circ}$  & $4.60 \pm 0.12$ & $-0.57 \pm 0.03$ & $-0.10 \pm 0.01$ \\
                                               & 3 & $180^{\circ}$  & $4.57 \pm 0.09$ & $-1.20 \pm 0.03$ &  $0.13 \pm 0.01$ \\
  {\large ${^{3}\!J}_{H^{N} C^{\beta}}$      } & 1 & $60^{\circ}$   & $2.72 \pm 0.03$ & $-0.35 \pm 0.03$ & $0.12 \pm 0.01$  \\
                                               & 2 & $60^{\circ}$   & $3.00 \pm 0.04$ & $-0.26 \pm 0.03$ & $-0.28 \pm 0.02$ \\
                                               & 3 & $60^{\circ}$   & $2.52 \pm 0.03$ & $-0.03 \pm 0.02$ &  $-0.09 \pm 0.02$ \\
  {\large ${^{3}\!J}_{H^{N} H^{\alpha}}$     } & 1 & $-60^{\circ}$  & $7.11 \pm 0.05$ & $-1.38 \pm 0.03$ & $1.43 \pm 0.04$  \\
                                               & 2 & $-60^{\circ}$  & $7.50 \pm 0.07$ & $-1.50 \pm 0.02$ & $1.50 \pm 0.06$  \\
                                               & 3 & $-60^{\circ}$  & $6.97 \pm 0.07$ & $-1.49 \pm 0.04$ &  $1.63 \pm 0.05$  \\
\hline
\end{tabularx}
    \begin{tablenotes}
      \item[1] 1D3Z as the structural ensemble
      \item[2] 2NR2 as the structural ensemble
      \item[3] RosettaFold2 (RF2) as the structural ensemble
    \end{tablenotes}
\end{threeparttable}
\end{table}

\begin{figure*}[!htb]
  \includegraphics[width=\linewidth]{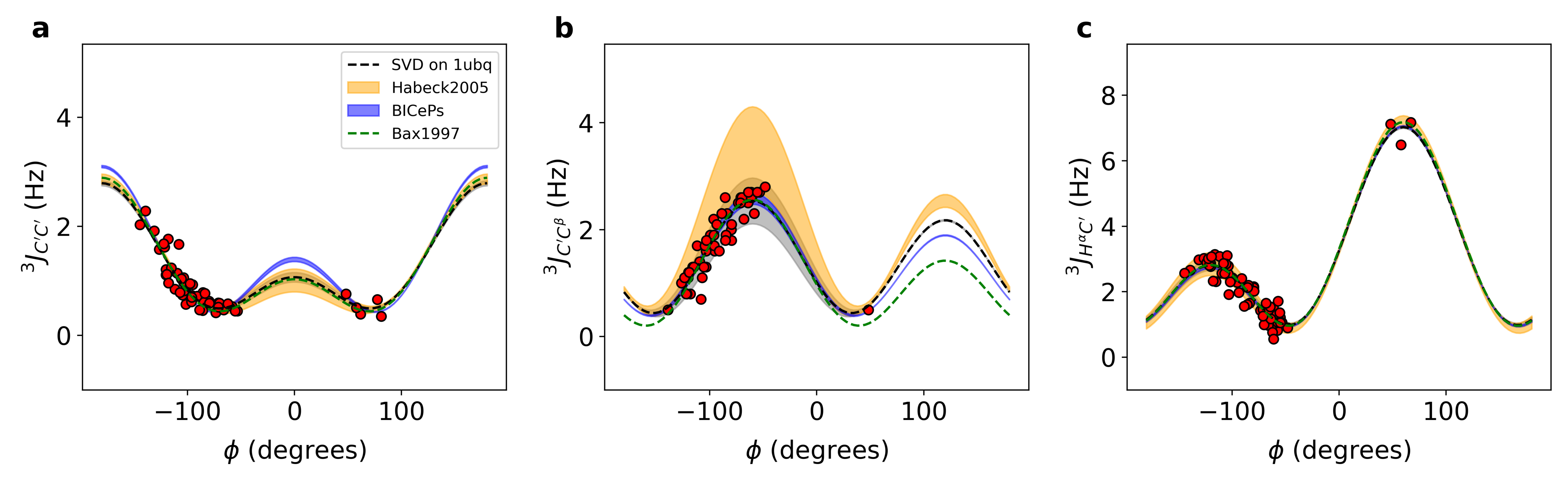}
  \caption{\small Karplus curves with BICePs-refined Karplus coefficients using the 1d3z ensemble for (a-c) ${^{3}\!J}_{C^{\prime}C^{\prime}}$, ${^{3}\!J}_{C^{\prime}C^{\beta}}$, and ${^{3}\!J}_{H^{\alpha} C^{\prime}}$. For comparison, SVD on 1ubq using experimental scalar coupling constants with $\phi$-angles derived from the X-ray structure (black dashed line), and red dots correspond to the fitted data points. Additionally, parameterizations from Bax et al. 1997 (green) and parameterization from Habeck et al. 2005 (yellow) were overlaid for comparison. The thickness of the line corresponds to the uncertainty.}
  \label{fig:half_karplus_curves}
\end{figure*}

Figure \ref{fig:half_karplus_curves} compares the Karplus curves derived from BICePs using the 1D3Z ensemble with previously published parameters obtained from NMR refinements, showing subtle differences. Both the marginal posterior distributions of the FM parameters and the Karplus curves for each scalar coupling demonstrate significant congruence with the historical NMR refinement results\cite{cornilescu1998validation,habeck2005bayesian}.  For all six types of $J$-coupling, see Figure \ref{fig:all_karplus_curves}.

The predicted parameters have large similarities across structural ensembles. BICePs-predicted coefficients using the 1D3Z ensemble (Figure \ref{fig:all_coefficient_histograms}) and predicted coefficients using the RF2 ensemble  (Figure \ref{fig:all_coefficient_histograms_RF2}) are found to strongly overlap. Furthermore, the traces of the FM parameters over time (Figure \ref{fig:all_coefficient_traces}) confirm convergence.

A key advantage of BICePs is its ability to sample the posterior densities of forward model uncertainties, $p(\sigma)$ (Figure \ref{fig:marginal_distributions_of_sigma}).  For certain sets of $J$-coupling constants (e.g., ${^{3}\!J}_{H^{N} H^{\alpha}}$ and  ${^{3}\!J}_{H^{N} C^{\prime}}$) the marginal posterior distribution of the variance scaling parameter $p(\varphi)$ has a sampled mean slightly larger than 1.0, indicating that the functional form of the likelihood opted for long tails to account for a few outlier data points deviating from the mean.

\begin{figure}[!htb]
  \includegraphics[width=\linewidth]{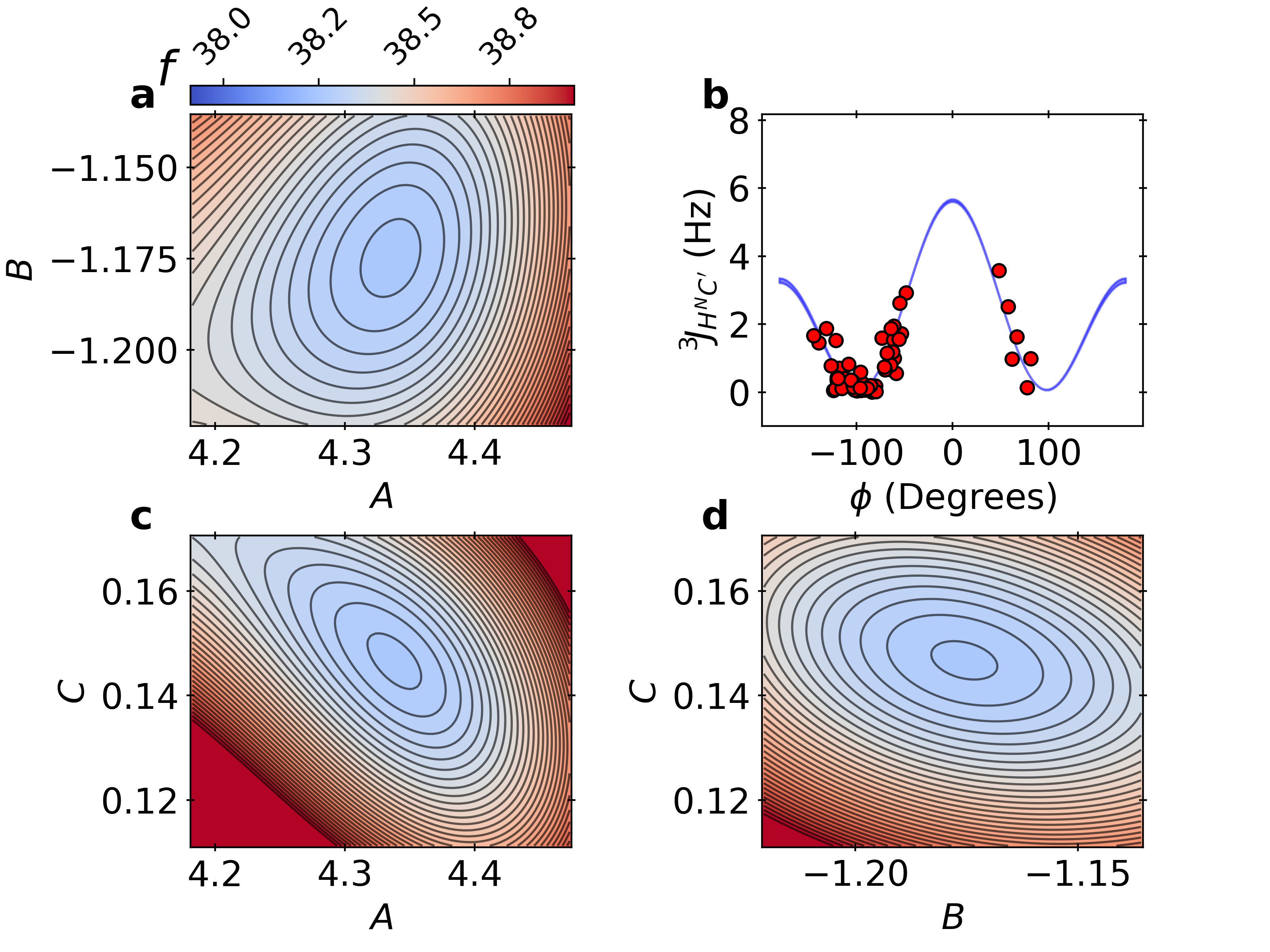}
  \caption{\small Landscapes of the BICePs score with respect to the predicted Karplus coefficients for ${^{3}\!J}_{H^{N} C^{\prime}}$. Panels a, c and d illustrate the energy landscape $f$ for pairs of Karplus coefficients when using the 1D3Z structural ensemble during refinement.}
  \label{fig:karplus_coefficent_landscape}
\end{figure}

\paragraph*{The BICePs free energy landscape for ${^{3}\!J}_{H^{N} C^{\prime}}$ Karplus parameters.} In Figure \ref{fig:karplus_coefficent_landscape}, we show the free energy landscape, which is also equivalent to the BICePs score landscape $f_{\xi=0 \rightarrow 1}$.  The Karplus relation for ${^{3}\!J}_{H^{N} C^{\prime}}$ was found to overlap strongly with the results obtained by SVD when using $\phi$ angles from the X-ray crystal structure (Figure \ref{fig:all_karplus_curves}).  Red data points are shown using the experimental $J$-couplings with $\phi$ angles derived from X-ray crystal pose 1UBQ\cite{vijay1987structure}.
The joint BICePs score landscape for the six sets of parameters is too complex to visualize. Instead, we constructed a smooth 2-D landscape for each pair of parameters within a set of scalar couplings by training a Gaussian process on the BICePs energy trace using a radial basis function (RBF) kernel with an additive white noise to account for observational uncertainty.  The characteristic length scale was bounded within $[0.1, 10.0]$ and the signal variance was set to 1.0 and the noise variance was set to $10^{-5}$.  The landscape matches the computed BICePs scores, and shows minima in the correct locations. All BICePs score landscapes for each of the six sets of Karplus coefficients are illustrated in Figure \ref{fig:all_landscapes}.

\begin{figure*}[!htb]
  \includegraphics[width=\linewidth]{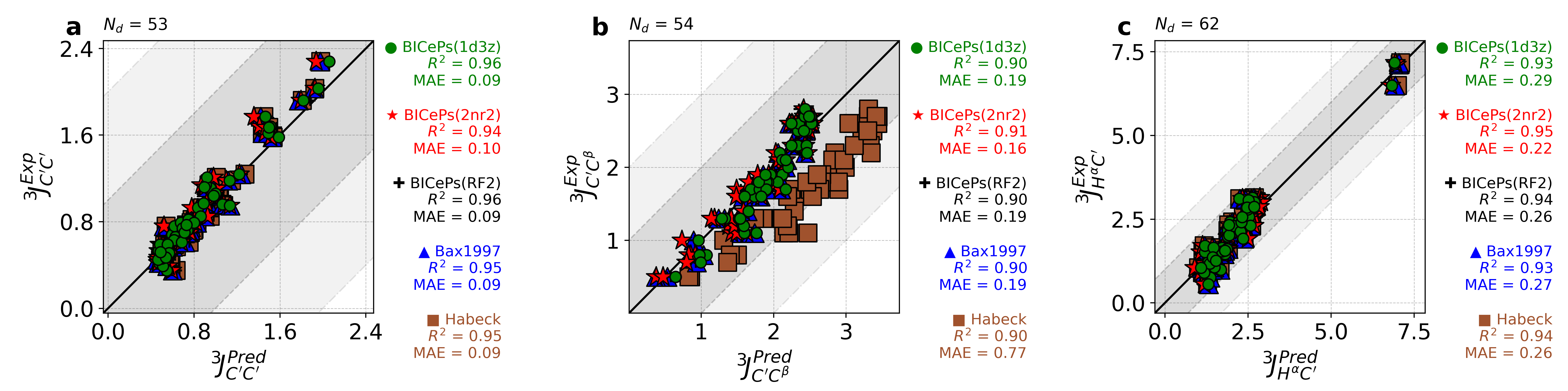}
  \caption{\small Validation of BICePs-predicted Karplus coefficients perform similarly to Bax1997 and achieve minor improvements over Habeck2005 for scalar coupling predictions for the simulated ensemble of CHARMM22*. Each panel for (a) ${^{3}\!J}_{H^{\alpha} C^{\prime}}$,  (b) ${^{3}\!J}_{C^{\prime}C^{\beta}}$, and (c) ${^{3}\!J}_{C^{\prime}C^{\prime}}$ shows strong correlations between predictions and experiment. Karplus coefficients derived from BICePs using the 2NR2 ensemble gives the best performance for CHARMM22*. For the remaining sets of $J$-coupling, please see Figure \ref{fig:correlations_charmm22*}.}
  \label{fig:correlation_charmm}
\end{figure*}

\begin{figure}[!htb]
  \includegraphics[width=0.9\linewidth]{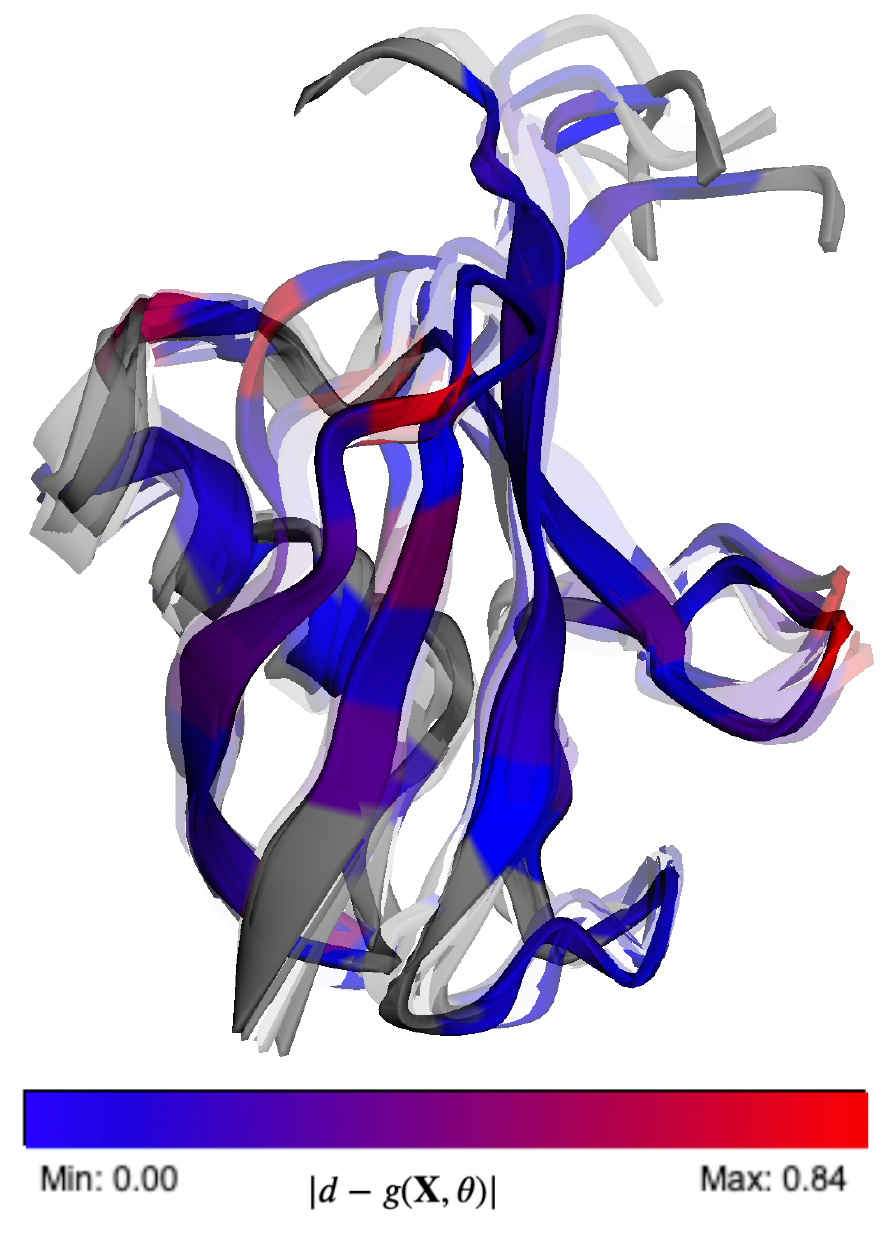}
  \caption{\small  A visualization of the 2NR2 conformational ensemble after joint conformational populations and Karplus parameter optimization using BICePs.   Structures (144 in total) with higher transparency have lower populations.  Residues are colored by the magnitude of the deviation between the replica-averaged forward model observations $g(\mathbf{X}, \theta)$ and the experimental ${^{3}\!J}_{C^{\prime}C^{\prime}}$ data $d$, where the color is normalized from blue (lowest deviation) to red (highest deviation).}
  \label{fig:2NR2_ensemble}
\end{figure}

To demonstrate the transferability across different generative models and validate our parameters, we evaluated the accuracy of the back-calculated scalar couplings using the different sets of Karplus coefficients.  In Figure \ref{fig:correlation_charmm}, we illustrate how the various sets of parameters derived from different techniques and different structural ensembles exhibit similar performance metrics.  Interestingly, applying BICePs-refined Karplus parameters to an ensemble generated by a molecular dynamics simulation (CHARMM22*),\cite{Piana2013}  some parameter sets are revealed to be more transferable than others. The mean absolute error (MAE) and coefficient of determination ($R^{2}$) for all six types of scalar couplings across different structural ensembles are shown in Figures \ref{fig:correlations_1d3z}-\ref{fig:correlations_charmm22*}.  On average, the BICePs-refined parameters derived from the 2NR2 ensemble (BICePs(2NR2)) give the lowest MAE between experiment and predictions for the CHARMM22* simulated ensemble, closely followed by BICePs(RF2) parameters, whereas Habeck 2005 has the highest due to known difficulties with ${^{3}\!J}_{C^{\prime}C^{\beta}}$. A visualization of the BICePs-reweighted 2NR2 structural ensemble using BICePs(2NR2) parameters is shown in Figure \ref{fig:2NR2_ensemble} with higher transparency representing lower populations. Residues are colored by the magnitude of the replica-averaged data deviation from the experiment in ${^{3}\!J}_{C^{\prime}C^{\prime}}$, where the color is normalized from blue (lowest deviation) to red (highest deviation).

To quantify which parameters produce the best predictions for ubiquitin, we compute BICePs scores, $f_{\xi=0 \rightarrow 1}$ for each of the structural ensembles. This score directly relates to the quality of FM parameters and their predictive accuracy at reproducing experimental scalar couplings, integrating over all sources of error. Lower BICePs scores indicate better agreement with experiment. BICePs scores were computed using eleven intermediate $\xi$-values between 0 and 1, which were optimized using \textit{pylambdaopt}.\cite{novack2025simple}  The results are shown in Table \ref{tab:biceps_scores}.  Each row in Table \ref{tab:biceps_scores} corresponds to BICePs scores using all six sets of Karplus coefficients used on different structural ensembles. The lowest score is shown in bold. (Note that the BICePs score is an extensive quantity that grows linearly with the number of replicas.  For this reason, our results report the \textit{reduced} BICePs score, $f(\theta)/N_r$. )

The leftmost column in Table \ref{tab:biceps_scores} corresponds to the parameters, where \textit{BICePs(1D3Z)} are the parameters in Table \ref{tab:karplus_parameters} (set 1), which used 1D3Z ensemble to obtain Karplus coefficients. BICePs score columns, e.g., $f_{\xi=0 \rightarrow 1}^{\text{1d3z}}$ corresponds to BICePs scores evaluated for the 1D3Z ensemble. That is, the superscript corresponds to the structural ensemble used as a validation step. The reported BICePs scores, were averaged over five independent rounds of validation each.  BICePs calculations burned for 1k MCMC steps, followed by 50k steps of production-run MCMC sampling.

The BICePs score, $f_{\xi=0 \rightarrow 1}^{\text{1d3z}}= 38.14 \pm 0.08$ (Table \ref{tab:biceps_scores}) is equivalent (within error) with the most probable landscape basin $f = 38.15 \pm  0.19$ from sampling the energy landscape, computed as an average across four chains; an example for one chain is shown in Figure \ref{fig:karplus_coefficent_landscape}.  This is additional evidence of the algorithm's reliability, and corroborates that the results from variational minimization of the BICePs score and full joint posterior sampling are equivalent.
\begin{table}[h]
  \caption{BICePs scores (32 replicas), $f$ for each structural ensemble over all sets of parameters, averaged over five independent rounds of validation each.}
\label{tab:biceps_scores}
\centering
\small
\begin{tabularx}{\linewidth}{lccc}
\hline
  Parameters & $f_{\xi=0 \rightarrow 1}^{\text{1d3z}}$ & $f_{\xi=0 \rightarrow 1}^{\text{2nr2}}$ & $f_{\xi=0 \rightarrow 1}^{\text{CHARMM22*}}$ \\
\hline
Bax 1997\cite{wang1996determination,cornilescu1998validation}      &  $ 61.12  \pm   0.08$ &  $132.49  \pm   0.09$ &  $ 99.27  \pm   1.91$ \\
Habeck 2005 \cite{habeck2005bayesian}  &  $135.66  \pm   0.08$ &  $199.32  \pm   0.24$ &  $165.64  \pm   0.47$ \\
SVD(1D3Z)    & $44.76 \pm 0.66$        &  $147.44 \pm 0.70$    & $120.63 \pm 0.21$ \\
BICePs(1D3Z) &  $\mathbf{38.14  \pm   0.08}$ &  $141.69  \pm   0.16$ &  $105.00  \pm   0.74$ \\
BICePs(2NR2) &  $118.42 \pm 0.14$ &  $\mathbf{113.15 \pm 1.34}$ &  $\mathbf{76.27 \pm 0.66}$ \\
BICePs(RF2)  &  $68.07 \pm 0.60$ &  $129.42 \pm 0.15$  &  $88.04 \pm 0.21$ \\
\hline
\end{tabularx}
\end{table}

For both the 2NR2 and CHARMM22* structural ensembles, Bax 1997, BICePs(RF2) and BICePs(1D3Z) parameters give very similar BICePs scores, which suggests robust accuracy of FM parameters in reproducing experimental scalar couplings and the transferability of FM parameters across different prior structural ensembles.  For the CHARMM22* simulated ensemble, the BICePs(2NR2) parameters give the lowest BICePs score. A partial explanation for this is that the structural ensemble 2NR2 deposited in the Protein Data Bank was generated using the CHARMM22 force field with additional experimental restraints during simulation.\cite{richter2007mumo}

We obtained Karplus parameters using SVD with uniform populations across the 1D3Z ensemble (Table \ref{tab:SI_svd(1d3z)_parameters}), denoted SVD(1D3Z). As shown in Table \ref{tab:biceps_scores}, these parameters perform well for 1D3Z itself ($f_{\xi=0 \rightarrow 1}^{\text{1d3z}} = 44.76 \pm 0.66$) but generalize poorly to other ensembles, giving higher scores for 2NR2 and CHARMM22*. The MAE results in Table \ref{tab:SI_MAE} show a similar trend: while SVD(1D3Z) predictions are competitive within 1D3Z, they are consistently worse than BICePs(1D3Z) for 2NR2 and CHARMM22*. This highlights a key limitation of SVD with uniform populations—without proper treatment of ensemble populations, the parameters risk overfitting to the training ensemble and lose transferability. By contrast, BICePs jointly optimizes both parameters and populations, yielding lower errors across ensembles and more robust performance.

It is difficult to say which of the model parameters are the best for ubiquitin, so we compare the top four: BICePs(RF2), Bax 1997, BICePs(1D3Z), and BICePs(2NR2).  The BICePs(2NR2) parameters are objectively better at predicting $J$-couplings from structures of ubiquitin generated from simulations using CHARMM22* force field.  In general, when looking across structural ensembles, the lowest BICePs scores come from the 1D3Z structural ensemble ($f_{\xi=0 \rightarrow 1}^{\text{1d3z}}$) except for BICePs parameters derived from the 2NR2 ensemble (BICePs(2NR2)). This confirms that the 1D3Z structural ensemble gives the strongest agreement with experimental NMR observations. However, our BICePs(RF2) parameters have the best transferability across structural ensembles and have lower BICePs scores compared to Bax1997 parameters.  In a separate study, we evaluated the performance of the BICePs(RF2) parameters (denoted as “Raddi2024” in ref\cite{NRV2024}) in predicting ${^{3}\!J}_{H^{N}H^{\alpha}}$ couplings for a set of non-natural and cyclic $\beta$-hairpin peptides\cite{NRV2024}. Our results show that these parameters yield consistently lower prediction errors compared to the Bax2007\cite{Bax2007} and Kessler1988\cite{kessler1988conformational} parameter sets. Based on all evidence, we recommend the BICePs(RF2) parameters for downstream applications such as structural refinement and validation. Nonetheless, users should carefully assess accuracy for their specific system, as additional refinement may still be required.

\section{Discussion}
\subsection*{Ensembles from deep generative models like RosettaFold2 can be used for parameter refinement.}
The booming field of machine learning and artificial intelligence is transforming the field of biomolecular modeling at a swift pace. Recent advancements in generative models, such as AlphaFold\cite{jumper2021highly}, RosettaFold\cite{baek2023efficient} and others, have heralded a new era in the accurate prediction of structural ensembles. Leveraging the predictive power of these models as structural priors is expected to help refine ensemble predictions when integrated with similar algorithms to BICePs\cite{brotzakis2025alphafold}. Here, we have demonstrated that structural ensembles generated from RosettaFold2 (RF2) can be reweighted to better align with experimental measurements, while simultaneously refining Karplus parameters. Validation of these parameters by the BICePs score and other statistics demonstrates improved accuracy across a variety of structural ensembles of ubiquitin.

\subsection*{Automatic determination of unknown errors.}
Our method provides a notable advantage by automatically estimating all potential error sources throughout the ensemble refinement process. This estimation is facilitated through the analysis of posterior distributions, which are instrumental in deriving accurate error assessments for the Karplus coefficients. Consequently, this negates the need for cross-validation techniques commonly used in other approaches\cite{frohlking2023simultaneous,vuister1993quantitative}.


In previous work, we have shown how the BICePs score is a better metric model validation than the traditional $\chi^{2}$ test.\cite{raddi2025model} Unlike $\chi^{2}$, which presupposes a fixed and known error, BICePs dynamically learns the complete posterior distribution of all uncertainty parameters, providing a measure of model quality that integrates over all sources of error. Our work here supports this premise.

%

\subsection*{Comparison of joint posterior sampling and variational minimization for FM parameterization.}
\label{sec:compare_methods}
Two strategies can be used for forward model parameter optimization in BICePs: joint posterior sampling and variational minimization. Joint posterior sampling, which we recommend, is efficient with modern gradient-based samplers such as Hamiltonian Monte Carlo (HMC), provides direct estimates of parameter uncertainties and covariances, and is fast in practice (50,000 steps in ~15 seconds on a MacBook M1 Pro). Variational minimization of the BICePs score can be advantageous when handling many parameters and is straightforward to adapt for different forward models, especially in convex landscapes, but it requires repeated free energy calculations via MBAR at each epoch, resulting in longer runtimes, and its uncertainty estimates are less reliable in non-convex landscapes. Overall, posterior sampling is faster and more robust, while variational minimization may be preferable for very high-dimensional, convex problems.

\subsection*{Bayesian ranking of Karplus-type relations}
Although here we optimize Karplus relations with a standard functional form, many possible relations can be used to accommodate the diverse characteristics of molecular structures, from rigid to flexible. \cite{minch1994orientational} The BICePs algorithm can determine coefficients and their uncertainties for any functional form, including those with additional parameters. Although we do pursue this aim in our current work, it is straightforward to use Bayesian model selection to objectively rank empirical models based on their BICePs scores, while automatically accounting for model complexity, thus providing a balance of model accuracy and parsimony.

\subsection*{Application to Non-Natural and Cyclic Peptides for fine-tuning folding landscapes.}
Our framework for forward model optimization has recently been successfully applied to a challenging test case involving a diverse set of non-natural and cyclic $\beta$-hairpin peptides\cite{NRV2024}. In that study, we used BICePs' forward model optimization approach to refine Karplus parameters for ${^{3}\!J}_{H^{N} H^{\alpha}}$ predictions directly against experimental NMR data. With an improved Karplus relation, BICePs reshaped the folding landscapes to better agree with NMR data, enabling more accurate predictions of folding stability for each peptide. This work highlights the resolution enhancement gained by the improved forward model predictions and population reweighting using BICePs, enabling the detection of subtle changes in folding stability—such as those introduced by a sidechain hydrogen- or halogen-bonding group, which alters stability by no more than 2 kJ/mol—offering a reliable pathway for designing foldable non-natural and cyclic peptides.

\subsection*{The BICePs score can be used as a loss function to train neural networks and other differentiable forward models.}
A central challenge in physical sciences is integrating domain knowledge with data-driven models. Traditional machine learning often lacks physical consistency or interpretability, while purely physics-based models may miss subtle experimental trends. One way to bridge this gap is to use the BICePs score—a free energy-based Bayesian loss—to train machine learning forward models using ensemble-averaged experimental data. This allows one to parameterize neural networks or other differentiable models in a physically grounded, uncertainty-aware fashion. In this way, BICePs provides a general-purpose learning framework for constructing interpretable, probabilistically-justified forward models for complex observables.  This capability may be especially valuable in contexts where observables are too complex to be modeled analytically—such as in chemical shift prediction or other nonlinear structure–observable relationships\cite{han2024accurate,fang2021predicting,sanchez2021deepemhancer}.

To demonstrate the viability of using BICePs for neural network training, we constructed a toy model for $J$-coupling prediction. In this model, the backbone dihedral angle $\phi$ was the sole input variable. To appropriately encode the periodicity of angular data, each $\phi$-angle was embedded into two dimensions via a sine–cosine transformation:
\begin{equation}
x = \begin{bmatrix}
\cos(\phi) \\
\sin(\phi)
\end{bmatrix}
\in \mathbb{R}^2
\end{equation}
This representation ensures continuity across the periodic boundary at $\pm \pi$ and provides a smooth input feature space for learning.

A fully connected feedforward neural network was constructed using this 2D input, consisting of one hidden layer with 200 neurons and Gaussian Error Linear Unit (GELU) activation functions. The final output was a scalar prediction corresponding to the $J$-coupling value. To ensure stable training dynamics compatible with GELU, the weights were initialized using LeCun normal initialization\cite{lecun2002efficient}.

The network was trained using ADAM optimization for 2000 epochs with a learning rate of $10^{-3}$, using the Good-Bad likelihood model with the BICePs score as the loss function. The toy dataset consisted of 5 conformational states, each with 60 synthetic $J$-coupling observables using the Karplus relation $J(\phi) = A \cos^2(\phi) + B \cos(\phi) + C$, with ground truth coefficients $A = 6.51$, $B = -1.76$, and $C = 1.60$. Our training dataset consisted of 300 $\phi$-angles as NN inputs.

First, we test the performance without adding random or systematic noise to the synthetic experimental J-coupling observables.  Figure \ref{fig:toy_NN_example}a shows the convergence behavior of the BICePs score across five independent training runs, where a plateau is reached around step 1500. Figure \ref{fig:toy_NN_example}b shows the Karplus curves predicted by the five trained networks, along with their mean and standard deviation (blue curve). For comparison, we also include the result of singular value decomposition (SVD) applied directly to the Karplus form under the assumption that backbone $\phi$ angles are known precisely, as would be the case with a high-resolution crystal structure. The parameters from SVD were determined to be $A = 6.47 \pm 0.002$, $B = -1.75 \pm 0.001$, and $C = 1.62 \pm 0.001$. Our framework yields comparable predictive performance, where the predictive accuracy of J-couplings for the SVD approach achieved an RMSE of $0.01 \pm 0.001$ Hz, while the BICePs-trained neural network reached an RMSE of $0.04 \pm 0.01$ Hz.

Next, we evaluate the performance when adding random or systematic noise to the synthetic experimental $J$-coupling observables ($\sigma_{\text{data}} = 0.79$ Hz).  Figure \ref{fig:toy_NN_example_rand_sys_error}a--b shows the convergence behavior of the BICePs score and gradient norm across five independent training runs, where convergence begins around 1000 epochs. The BICePs score was determined to be $f_{\xi=0\rightarrow 1} = 62.29$ nats. Figure~\ref{fig:toy_NN_example_rand_sys_error}c shows the Karplus curves predicted by the five trained networks, along with their mean and standard deviation (blue curve).  Comparison against SVD illustrates that BICePs, when used to train neural networks from conformational ensembles alone, recovers predictive performance comparable to analytical fitting based on ground-truth dihedral angles, even in the presence of random and systematic errors. When comparing predictive accuracy of $J$-couplings, the SVD approach achieved an RMSE of $0.24 \pm 0.03$ Hz, while the BICePs-trained neural network reached an RMSE of $0.21 \pm 0.01$ Hz, demonstrating that our framework yields improved predictive performance. For completeness, we also used BICePs to optimize the parameters of the Karplus relation in this example, obtaining $A = 6.6 \pm 0.05$, $B = -1.7 \pm 0.023$, and $C = 1.7 \pm 0.037$, which resulted in the lowest RMSE of $0.17 \pm 0.0001$ Hz. This outcome is expected, since the “true” values were originally generated using the Karplus relation. By contrast, the neural network achieves remarkably low error without being constrained to any functional form, highlighting its flexibility and strong predictive capability.

Convergence of the BICePs score is typically accompanied by a reduction in the posterior uncertainty parameter $\sigma$, which sharpens the likelihood landscape and amplifies gradients with respect to model parameters. Specifically, the derivative of the BICePs score with respect to a parameter $\theta$ contains a term proportional to $1/\sigma^2$ (see equations~\ref{eq:GB_derivative} and~\ref{eq:Gaussian_derivative}), meaning that as $\sigma \to 0$ the sensitivity of the loss to prediction errors increases.

During training of the NN-based forward model, the posterior sampling of  $\sigma$  plays a critical role in regularization. If $\sigma$ were held fixed during training, the model would be limited in its ability to adjust the sharpness of the likelihood surface and could fail to reach an optimal fit to the data. By treating $\sigma$ as a nuisance parameter and marginalizing over it within the BICePs framework, the model adaptively tunes the effective strength of the loss function, yielding an implicit form of regularization. This dynamic adjustment helps stabilize training and ensures that neural network parameters are optimized in a manner consistent with both the experimental data and the underlying physical ensemble.

Despite these advantages, neural networks trained on limited or biased datasets can still face overfitting risks. Several mitigation strategies can be incorporated within the BICePs training framework: (1) data augmentation via additional synthetic observables or sampling of underrepresented dihedral regions, (2) regularization through dropout layers to discourage over-parameterization, and (3) physically informed constraints on network architecture or inputs, such as torsional symmetries—as we did for J-coupling or local environment descriptors for chemical shifts. These strategies, combined with the adaptive regularization provided by marginalizing over $\sigma$, further safeguard against overfitting and improve model generalizability.

While this toy model serves as a useful proof-of-concept, the broader applicability of our framework is in modeling more complex observables such as NMR chemical shifts. These observables are typically nonlinear and high-dimensional functions of atomic coordinates, making them difficult to model analytically. In contrast, neural networks trained using BICePs can serve as flexible surrogates for such forward models. Importantly, this framework supports the simultaneous training of multiple neural networks, each targeting a distinct observable—for example, one network may predict $^3J_{\text{HN–H}\alpha}$ couplings, another $^3J_{\text{C}’–C’}$ couplings, and others may be used for predicting chemical shifts. In the case of backbone chemical shift prediction, a separate neural network is generally constructed for each nucleus type (e.g., C$\beta$, $^1$H$\alpha$, and $^1$HN), enabling specialized learning tailored to the distinct structural dependencies of each observable.  All networks can be trained jointly using a shared conformational ensemble, with BICePs enforcing consistency with the corresponding experimental data.

Models have limited predictive reliability in regions lacking sufficient training data. A central advantage of our framework lies in its ability to incorporate physically relevant information through the choice of input features or explicit regularization. In our example using the toy model system, we employed the backbone dihedral angle $\phi$ as the sole input feature (projected to the unit circle to maintain periodicity). However, neural network inputs can encompass a broader range of features for complex observables such as chemical shifts, including quantum calculations, partial charges, descriptors of local atomic environments, interatomic distances, hydrogen bonding patterns, solvent accessibility, or torsional strain energies. These diverse features encode crucial structural and electronic contexts and can be precomputed from molecular simulations or quantum calculations before training.

We did not impose physical constraints or priors when training our model to predict $J$-coupling constants. Users might explicitly incorporate such physical priors or constraints through architectural designs, regularization terms, or physically-informed input features, especially critical when dealing with sparse or unevenly distributed training data. For instance, in our experiments, neural networks employing Rectified Linear Unit (ReLU) activations struggled to accurately predict $J$-couplings at backbone dihedral angles greater than $+90^\circ$. In contrast, GELU activations provided smoother and more reliable extrapolation in these underrepresented regions. Thus, choosing suitable activation functions and rich input features improves predictive accuracy across the conformational landscape.

\subsection*{Relation to mixture-based outlier models and multi-level outlier severity.}
Our outlier-aware Good-and-Bad likelihood model uses a two-component mixture $M=2$ of variances. One down-weights data affected by systematic error, while the other preserves the influence of "good" data. A natural extension of the Good-Bad model replaces a single "bad" mode with multiple levels of outlier severity.

In reliability and survival analysis, censored finite-mixture models help separate heterogeneous data while accounting for censoring. For example, Akhtar \& Alharthi\cite{akhtar2025bayesian} apply a censored two-component mixture of geometric distributions under a Bayesian framework, illustrating how finite mixtures capture central trends while disregarding extreme values.
Although distinct from our context, it illustrates the broader principle that finite mixtures
model structured heterogeneity rather than ad hoc data exclusion.

For a natural extension to our Good-Bad model, we assign $p(\sigma_{j} | \sigma_{0})$ from equation \ref{eq:posterior_marginal_over_sigma} to be a multi-severity Good–Bad prior
\begin{equation}
p(\sigma_j\mid\sigma_0,\{\omega_m\},\{\varphi_m\})
=\sum_{m=1}^{M}\omega_m\,\delta\big(\sigma_j-\varphi_m\sigma_0\big),
\label{eq:GB_prior_multi_severity}
\end{equation}
where $\omega_m\ge 0$ are mixture weights with $\sum_{m=1}^{M}\omega_m=1$, and the severity inflators satisfy $\varphi_1\equiv 1 \lt \varphi_2 \lt \cdots \lt \varphi_M$. Thus, $\varphi_m$ scales the typical uncertainty $\sigma_0$ to represent increasingly severe outliers (mild $\to$ severe). The standard two-component Good–Bad model is recovered at $M=2$.

Inserting equation \ref{eq:GB_prior_multi_severity} into the integrand of equation \ref{eq:posterior_marginal_over_sigma} gives, for each observable $j$,
\begin{equation}
\begin{aligned}
\int\limits_{\sigma^{\mathrm{SEM}}}^{\infty}&\mathcal N\big(d_j\mid g_j,\sigma_j^2\big)\,
\sum_{m=1}^{M}\omega_m\,\delta(\sigma_j-\varphi_m\sigma_0)\,d\sigma_j \\
&=\sum_{m=1}^{M}\omega_m\,H\big(\varphi_m\sigma_0-\sigma^{\mathrm{SEM}}\big)\,
\mathcal N\big(d_j\mid g_j, \varphi_m^2\sigma_0^2\big),
\end{aligned}
\end{equation}
i.e., each observation is evaluated under a finite Gaussian scale mixture with the Heaviside factor enforcing the $\sigma^{\mathrm{SEM}}$ lower bound.

If we integrate out the mixing weights ${\omega_m}$ as in our two-component case, a uniform Dirichlet prior that is flat on the simplex ${\omega_m\ge0,\ \sum_m \omega_m=1}$ implies $\mathbb{E}[\omega_m]=1/M$ by symmetry, leading to
\begin{equation}
\begin{aligned}
  \sum_{m=1}^{M} & \left[ \int\limits_{\Delta^{M-1}}\omega_m\,p(\boldsymbol{\omega})\,d\boldsymbol{\omega} \right] \,   H\big(\varphi_m\sigma_0-\sigma^{\mathrm{SEM}}\big) \mathcal N\big(d_j\mid g_j, \varphi_m^2\sigma_0^2\big) \\
&=\frac{1}{M}\sum_{m=1}^{M}
H\big(\varphi_m\sigma_0-\sigma^{\mathrm{SEM}}\big)\,
\mathcal N\big(d_j\mid g_j,\ \varphi_m^2\sigma_0^2\big).
\end{aligned}
\end{equation}

Practically, this multi-severity formulation leaves a single Bayesian uncertainty parameter $\sigma_{0}^{B}$ and introduces $M-1$ free severity parameters ${\varphi_m}$ (with $\varphi_1=1$ fixed) to be inferred in the posterior. This extends the single-severity Good–Bad model used in our framework (where only a single $\varphi$ is sampled alongside $\sigma_{0}^{B}$) and provides finer control over how aggressively outliers are down-weighted.

\section{Conclusion}

In the quest for accurate forward model predictions, specifically for $J$-coupling, researchers often navigate the vast literature seeking Karplus parameters that align with their specific systems, occasionally settling for less-than-ideal solutions. Our work demonstrates that BICePs can be used as a robust tool for determining forward model (FM) parameters by sampling over the full posterior distribution of their values and uncertainties.

We used a toy model of protein dihedral angles to demonstrate that posterior sampling and variational minimization of the BICePs score are valid and robust approaches for FM parameter refinement. Using structural ensembles and experimental data for ubiquitin, we applied BICePs with posterior sampling to optimize six different sets of Karplus coefficients using different types of $J$-coupling measurements, while effectively addressing both random and systematic errors.

From these results, one can see how BICePs can be applied more generally to optimize a wide variety of forward models, including those represented by neural networks. By treating the BICePs score as a physically grounded loss function, our framework enables the training of data-driven forward models that are consistent with ensemble-averaged experimental observables and incorporate principled uncertainty quantification. These advances not only contribute to the refinement of molecular simulations but also provide a scalable and extensible machine learning approach for physical sciences, with applications in structural dynamics, model validation, and the construction of predictive, interpretable models across a broad spectrum of data regimes.

\section*{Conflicts of interest}
Authors declare no conflicts of interest.

\section*{Data and Software}
The BICePs algorithm is openly available at \href{https://github.com/vvoelz/biceps}{github.com/vvoelz/biceps}. All calculations in this work were performed using the development version biceps\_v3.0a, which is planned to be merged into the main branch. The specific version can be accessed at: \href{https://github.com/vvoelz/biceps/tree/biceps_v3.0a}{github.com/vvoelz/biceps/tree/biceps\_v3.0a}.
All scripts, Jupyter notebook examples and analysis can be found here. Detailed documentation and tutorials can be found here: \href{https://biceps.readthedocs.io}{biceps.readthedocs.io}. For any issues or questions, please submit the request on GitHub.

\section*{Acknowledgements}
The authors acknowledge that substantial content in this manuscript has also been included in the doctoral dissertation of RMR.\cite{raddi2024thesis} RMR, TM and VAV are supported by National Institutes of Health grant R01GM123296. This research includes calculations carried out on HPC resources supported in part by the National Science Foundation through major research instrumentation grant number 1625061 and by the US Army Research Laboratory under contract number W911NF-16-2-0189.
The authors also thank Kim Sharp for valuable discussions on extending the Good-and-Bad likelihood framework to incorporate multiple modes of data severity, and Somaiyeh Dadashi for her assistance with illustrations.

\bibliography{references}

\newpage

\section*{TOC Graphic}

\begin{figure}[htb!]
\centering
  \includegraphics[width=0.75\linewidth]{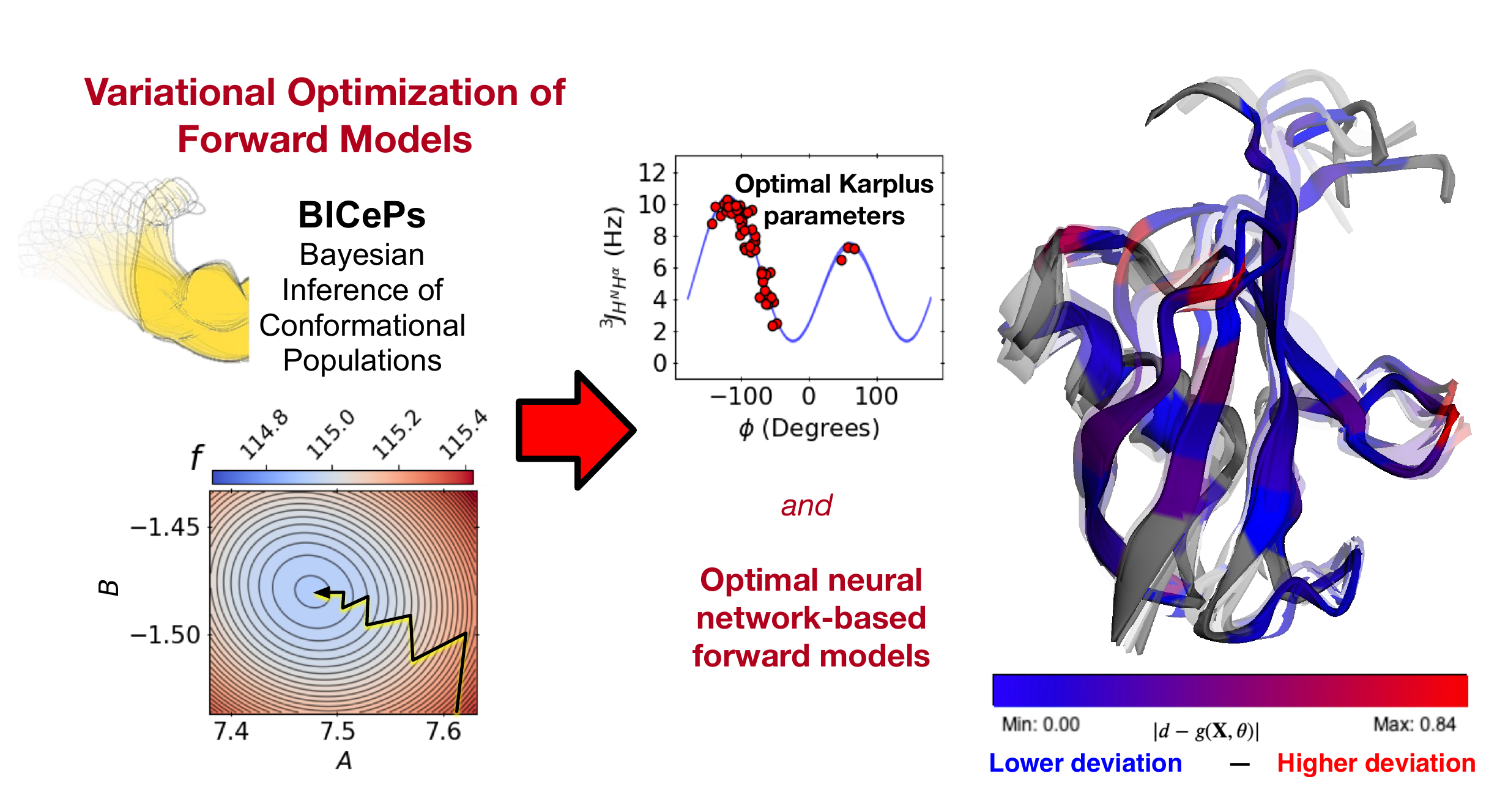}
\end{figure}

\newpage

\setcounter{figure}{0}
\renewcommand{\thefigure}{S\arabic{figure}}
\makeatletter
\renewcommand{\theHfigure}{Supp.\arabic{figure}} 
\makeatother

\setcounter{table}{0}
\renewcommand{\thetable}{S\arabic{table}}
\makeatletter
\renewcommand{\theHtable}{Supp.\arabic{table}} 
\makeatother

\setcounter{equation}{0}
\renewcommand{\theequation}{S\arabic{equation}}
\makeatletter
\renewcommand{\theHequation}{Supp.\arabic{equation}} 
\makeatother

\clearpage
\pagenumbering{arabic}
\setcounter{page}{1}
\renewcommand{\thepage}{S\arabic{page}}

{
\onecolumngrid

\setlength{\parskip}{\baselineskip}

\begin{center}

\noindent{{\Large Supplemental Information}}

\noindent{{\Large Automatic Forward Model Parameterization with \\Bayesian Inference of Conformational Populations}}

\noindent{{\large Robert M. Raddi, Tim Marshall and Vincent A. Voelz}}

\vspace{50px}
\end{center}

}

\twocolumngrid

\section*{Appendix A: Second partial derivatives of the BICePs score with respect to forward model parameters.}\label{sec:SI_theory}
The second partial derivatives of the BICePs score with respect to the FM parameters can be used for second-order optimization methods and uncertainty quantification.
For simplicity, we refrain from showing the complicated second derivatives for the Good-Bad model in the general case. In the case when when  $\varphi=1$, however,  the second partial derivatives of the BICePs energy function with respect to parameters $\theta_a$ and $\theta_b$ are:
\begin{equation}
  \begin{aligned}
    \frac{\partial^{2} u}{\partial \theta_{m}\partial \theta_{n}}  = N \left[ \sum_{j=1}^{N_{d}}  - \frac{\partial^{2} g_{j}(X,\theta)}{\partial \theta_{m}\partial \theta_{n}} \frac{ (d_{j}-g_{j}(\mathbf{X}, \theta))}{ \sigma_{j}^2} \right. \\
    \left. + \frac{\partial g_{j}(X,\theta)}{\partial \theta_{m}} \cdot \frac{\partial g_{j}(X,\theta)}{\partial \theta_{n}}  \frac{1}{ {\sigma_j}^2}  \right] .
  \end{aligned}
  \label{eq:second_derivative_off_diag}
\end{equation}
When $m$ = $n$, the second partial derivative of the energy is just:
\begin{equation}
  \begin{aligned}
    \frac{\partial^{2} u}{\partial \theta_{m}^{2}}  = N \left[ \sum_{j=1}^{N_{d}} \left( - \frac{\partial^{2} g_{j}(X,\theta)}{\partial \theta_{m}^{2}} \frac{ (d_{j}-g_{j}(\mathbf{X}, \theta))}{ \sigma_{j}^2} \right. \right. \\
    \left. \left. + \left(\frac{\partial g_{j}(X,\theta)}{\partial \theta_{m}} \right)^{2} \frac{1}{ {\sigma_j}^2} \right) \right].
  \end{aligned}
  \label{eq:second_derivative}
\end{equation}

The second partial derivatives of the BICePs score with respect to parameters $\theta_m$ and $\theta_n$ are:
\begin{equation}
  \begin{aligned}
    \frac{\partial^{2} f(\theta)}{\partial \theta_{m} \partial \theta_{n}} = \bigg\langle \frac{\partial^{2} u }{\partial \theta_{m}\partial \theta_{n}} \bigg\rangle
    - \biggl(& \bigg\langle  \frac{\partial u }{\partial \theta_{m}} \cdot \frac{\partial u }{\partial \theta_{n}} \bigg\rangle -\bigg\langle  \frac{\partial u }{\partial \theta_{m}} \cdot \frac{\partial u }{\partial \theta_{n}} \bigg\rangle^{\prime} \biggr),
  \end{aligned}
\end{equation}
where the notation $\langle \cdot \rangle^{\prime}$ denotes the ensemble average with respect to $\left(\frac{1}{Z}\exp(-u)\right)^{2}$. The first term on the right is the ensemble-average second derivative of the energy function $u$ with respect to parameters $\theta_{m}$ and $\theta_{n}$ given in equation \ref{eq:second_derivative} (when $\varphi = 1$).

When $m$ = $n$, the second partial derivative of the BICePs score reduces to the difference between the ensemble-averaged second derivative of the energy $u$ and the variance of its first partial derivative:
\begin{equation}
  \begin{split}
    \frac{\partial^{2} f}{\partial \theta^{2}} &= \bigg\langle \frac{\partial^{2} u }{\partial \theta^{2}} \bigg\rangle
    - \left( \bigg\langle \left( \frac{\partial u }{\partial \theta}\right)^{2} \bigg\rangle
    -\bigg\langle \frac{\partial u }{\partial \theta} \bigg\rangle^{2} \right)
  \end{split}
  \label{eq:d2f}
\end{equation}
In practice, this calculation is performed using the MBAR free energy estimator for the BICePs score and it's derivatives, by sampling at several intermediates  $\xi= 0\rightarrow 1$, which enables accurate estimates of all quantities.

\section*{Appendix B: Method equivalence between joint posterior sampling and variational minimization.}\label{sec:SI_method_equiv}
\subsection*{Equivalence of objectives.}
We define the BICePs score in the main text (Eq. \ref{eq:biceps_score}), but rewrite it here for clarity. Let $Z(\theta)$ be the model evidence given a set of parameters against a well-defined constant reference $Z_{0}$, then
\begin{equation}
  f(\theta) = - \ln \left(Z(\theta) \big/ Z_{0}\right) ,
\end{equation}
where
\begin{equation}
  Z(\theta) = \iint \exp\left(-u(\mathbf{X}, \mathbf{\sigma}; \theta)\right) d \mathbf{X} d \mathbf{\sigma},
\end{equation}
and the unnormalized log-density
\begin{equation}
u(X,\sigma;\theta)\ =\ -\ln\!\big(   p(D | X,\sigma, \theta)p(X)p(\sigma)   \big) + C,
\end{equation}
for some constant $C$.

The joint posterior (Eq. \ref{eq:joint_posterior}) can be written as
\begin{equation}
  p(X, \sigma, \theta | D) \propto\ p(\theta)\,\exp\!\big(-u(X,\sigma; \theta)\big),
\end{equation}
Marginalizing $(X,\sigma)$ yields
\begin{equation}
\begin{aligned}
  p(\theta\mid D)\ &\propto\ p(\theta)\,Z(\theta) \\
  \implies -\ln p(\theta\mid D)\ &=\ f(\theta)\ -\ \ln p(\theta)\ + C.
\end{aligned}
\end{equation}
Therefore, maximizing $p(\theta|D)$ during joint posterior sampling is equivalent to minimizing the BICePs Score $f(\theta)$ (when applying a flat prior for $p(\theta)$).

\subsection*{Equivalence of gradients.}
Rewritting Eq. \ref{eq:biceps_score_gradient} as
\begin{equation}
\begin{split}
  \frac{\partial f(\theta)}{\partial \theta_{i}} &= \iint \frac{1}{Z(\theta)}  \left[ \frac{\partial u}{\partial \theta_{i}} \right] \exp \left( - u \right)  d \mathbf{X} d \mathbf{\sigma} = \bigg\langle \frac{\partial u}{\partial \theta_{i}} \bigg\rangle
\end{split}
\end{equation}

The joint posterior sampling approach can use $M$ uncorrelated MCMC samples $(X^{(m)},\sigma^{(m)})$ to approximate the first derivative of the BICePs score:
\begin{equation}
  \frac{\partial f(\theta)}{\partial \theta_i}\ \approx \frac{1}{M}\sum_{m=1}^{M} \frac{\partial u(X^{(m)},\sigma^{(m)})}{\partial \theta_{i}}
  \label{eq:average_over_derivatives}
\end{equation}

\section*{Methods}

\subsection*{Structural ensembles of human ubiquitin}
\paragraph*{1D3Z}
This structural ensemble consists of the 10 conformations deposited in the Protein Data Bank (PDB: 1D3Z) from the NMR structural refinement performed by Cornilescu et al.   \cite{cornilescu1998validation}. The ensemble was calculated by the program X-PLOR using 2727 NOE distance restraints and 98 dihedral angle restraints derived from homo- and heteronuclear \textit{J} couplings.

\paragraph*{2NR2}
The 2NR2 structural ensemble (144 conformations) was taken from Richter et al. (PDB: 2NR2), where the refinement was performed using the MUMO (minimal under-restraining minimal over-restraining) method \cite{richter2007mumo}. In this approach, simulations were started from the X-ray crystal pose\cite{vijay1987structure} in the presence of replica-averaged restraints to NOEs distances and $S^{2}$ Lipari–Szabo order parameters.  Simulations used TIP3P solvent and the CHARM22 force field, with an augmented potential energy function $E_{\text{total}} = E_{\text{CHARMM22}} + E_{\text{restraints}}$.  Scalar couplings were not used during the refinement, but were only used as a validation metric.

\paragraph*{RosettaFold2 (RF2)}
The RF2 structural ensemble was generated using RosettaFold2\cite{baek2023efficient} made available through a Colabfold notebook.\cite{mirdita2022colabfold} Default parameters were used, and 25 conformations were generated.

\paragraph*{CHARMM22*}
The CHARMM22* structural ensemble was derived from a Markov State Model (MSMs) we constructed from a one-millisecond simulation of ubiquitin's native state at 300 K from Piana et al.\cite{Piana2013}.  The PyEMMA Python package\cite{wehmeyerintroduction} was used to determine appropriate backbone featurizations using the Variational Approach for Markov Processes (VAMP) scoring function VAMP-2\cite{Wu2020}. Based on the VAMP-2 scores, inverse distances were selected as features, while torsions were excluded due to lower VAMP-2 scores and their minimal contribution when paired with inverse distances. When comparing distances and inverse distances, the similarity of the average scores and relative standard deviations show that both featurizations are adequate, and inverse distances was selected purely on the higher average scores. Time-lagged independent component analysis (tICA) followed by $k$-means clustering was used to partition the trajectory data into 500 discrete conformational states for MSM construction \cite{Perez-Hernandez2013,MacQueen1967}. For the BICePs calculation, each of the 500 states was given equal statistical weight to enforce the uniform prior $p(X)$.

\subsection*{Details of parameters used in posterior sampling}
Uncertainty parameters $\sigma$ were sampled on a grid of logarithmically-spaced values between 0.001 to 100, to enforce the Jeffrey's prior.   Each grid value in the list was a factor of 1.02 larger than the next: [\texttt{1.00e-03, 1.02e-03, 1.04e-03, 1.06e-03, ... 9.72e+01, 9.92e+01}], resulting in a list of 582 values. For the Good-Bad model, sampling of the extra nuisance parameter, $\varphi$ took place on a grid from 1 to 100 with 1000 equally-spaced points.

\subsection*{Details of $\xi$ optimization}
Optimization was performed for a maximum of 2M steps with a tolerance of $10^{-7}$ and $\alpha = 10^{-5}$. With increasing amounts of data restraint energy, the optimization problem becomes  more complicated and more iterations are required to converge. In the case of insufficent iterations, the $\xi$-optimization might return negative $\xi$-values, which is incorrect and not physical. Initially, we start with 11 $\xi$-values $\texttt{\{0.0, 0.1, ... 0.9, 1.0\}}$ and with shift to lower values of $\xi$ after optimization e.g., \texttt{\{0.0, 0.02, 0.04, 0.07, 0.12, 0.26, 0.40, 0.60, 0.90, 0.97, 1.0\}}.

\newpage

\begin{table}[!htb]
\begin{threeparttable}
  \caption{Coefficients for the Karplus relation $^{3}\!J(\phi) = A \cos^{2}(\phi + \phi_{0}) + B \cos (\phi + \phi_{0}) + C$, determined by SVD on the 1D3Z ensemble using uniform populations.}
  \label{tab:SI_svd(1d3z)_parameters}
\centering
\small
\begin{tabularx}{\linewidth}{lcrrrr}
\hline
  & $\phi_{0}$ & $\mathrm{A}$ (Hz) & $\mathrm{B}$ (Hz) & $\mathrm{C}$ (Hz) \\
\hline
  {\large ${^{3}\!J}_{C^{\prime}C}$          } & $0^{\circ}$    & $0.55 \pm 0.29$  & $-0.94 \pm 0.04$   & $ 0.740 \pm  0.04$  \\
  {\large ${^{3}\!J}_{C^{\prime}C^{\beta}}$  } & $60^{\circ}$   & $2.32 \pm 0.73$  & $ 0.12 \pm 1.07$   & $ 0.13 \pm  0.36$   \\
  {\large ${^{3}\!J}_{H^{\alpha} C^{\prime}}$} & $120^{\circ}$  & $4.23 \pm 0.07$  & $-2.24 \pm 0.09$   & $ 1.03 \pm  0.05$   \\
  {\large ${^{3}\!J}_{H^{N} C^{\prime}}$     } & $180^{\circ}$  & $4.87 \pm 0.40$  & $-1.36 \pm 0.10$   & $ 0.05 \pm  0.05$   \\
  {\large ${^{3}\!J}_{H^{N} C^{\beta}}$      } & $60^{\circ}$   & $3.04 \pm 0.12$  & $-0.21 \pm 0.14$   & $-0.35 \pm  0.05$   \\
  {\large ${^{3}\!J}_{H^{N} H^{\alpha}}$     } & $-60^{\circ}$  & $7.93 \pm 0.16$  & $-1.78 \pm 0.07$   & $ 0.85 \pm  0.14$   \\
\hline
\end{tabularx}
\end{threeparttable}
\end{table}

\begin{table}[h]
  \begin{minipage}{\textwidth}
    \caption{Mean absolute error (MAE) for SVD(1D3Z) and BICePs(1D3Z) parameter sets across three different ensembles. These MAE values for BICePs(1D3Z) are found in Figures \ref{fig:correlations_1d3z}-\ref{fig:correlations_charmm22*}.}
    \label{tab:SI_MAE}
    \centering
    \small
      \begin{tabular*}{\textwidth}{@{\extracolsep{\fill}} |l|c c|c c|c c|}
    \hline
     & \multicolumn{2}{c|}{1D3Z} & \multicolumn{2}{c|}{2NR2} & \multicolumn{2}{c|}{CHARMM22*} \\
     \cline{2-7}
     & SVD(1D3Z) & BICePs(1D3Z) & SVD(1D3Z) & BICePs(1D3Z) & SVD(1D3Z) & BICePs(1D3Z) \\
    \hline
    {\large ${^{3}\!J}_{C^{\prime}C}$}           & 0.11  & 0.11           & 0.17  & \textbf{0.16} & 0.10  & \textbf{0.09} \\
    {\large ${^{3}\!J}_{C^{\prime}C^{\beta}}$}   & 0.15  & \textbf{0.14}  & 0.22  & 0.22           & 0.20  & \textbf{0.19} \\
    {\large ${^{3}\!J}_{H^{\alpha} C^{\prime}}$} & 0.23  & 0.23           & 0.27  & 0.27           & 0.29  & 0.29          \\
    {\large ${^{3}\!J}_{H^{N} C^{\prime}}$}      & 0.29  & 0.29           & 0.35  & 0.35           & 0.34  & \textbf{0.33} \\
    {\large ${^{3}\!J}_{H^{N} C^{\beta}}$}       & 0.19  & 0.19           & 0.24  & 0.24           & 0.25  & 0.25          \\
    {\large ${^{3}\!J}_{H^{N} H^{\alpha}}$}      & 0.41  & 0.41           & 0.62  & \textbf{0.61}  & 0.72  & \textbf{0.70} \\
    \hline
  \end{tabular*}
  \end{minipage}
\end{table}

\begin{figure*}
\centering
  \includegraphics[width=\linewidth]{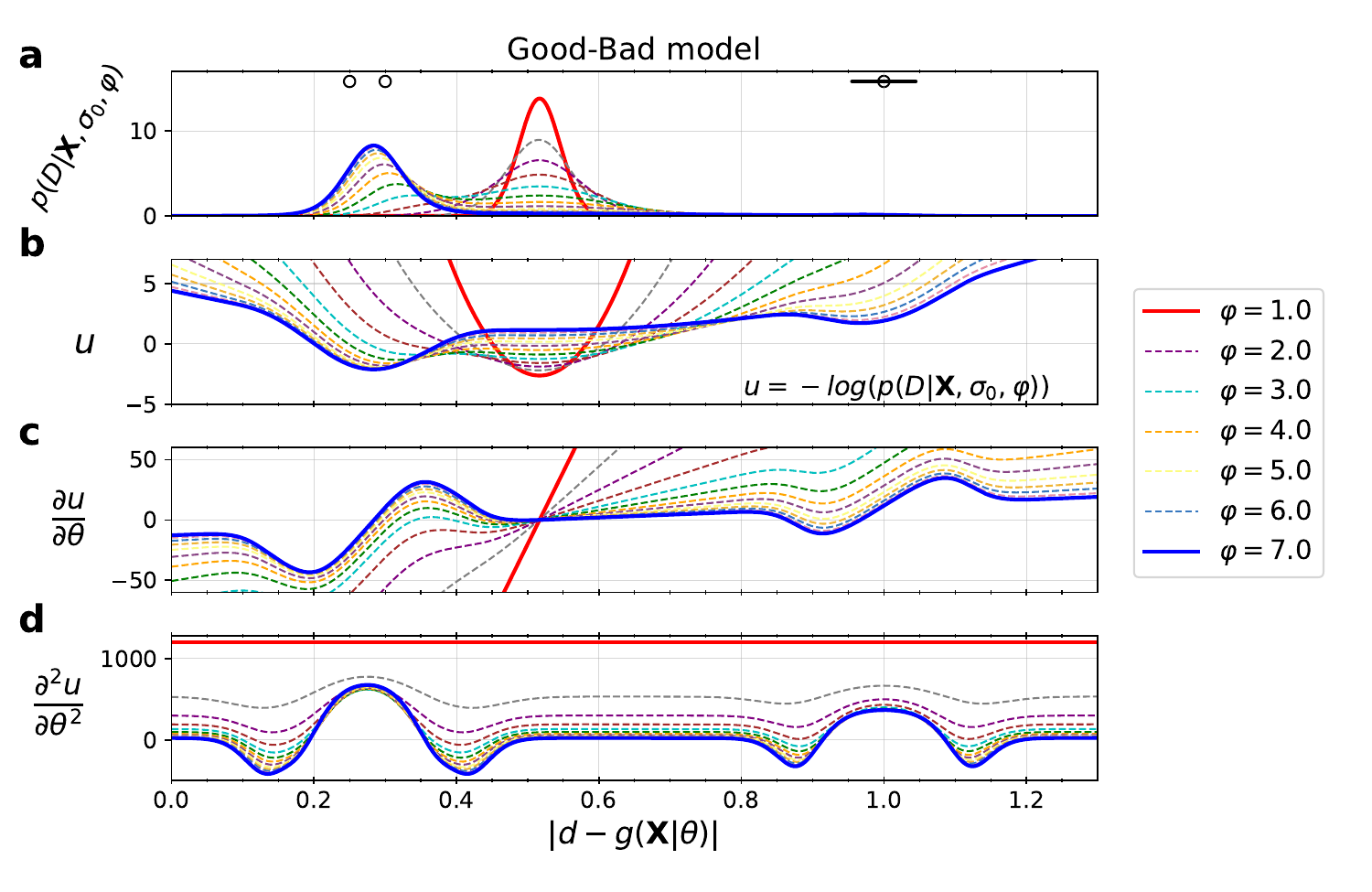}
  \caption{\small The Good-Bad model properly detects outliers. The probability density function (a) computed as $p(D|\mathbf{X}, \sigma_{0},\varphi) = \prod^{N_{j}}_{j} p(d_{j}|\mathbf{X}, \sigma_{0},\varphi)$ and energy landscape (b) of the marginal likelihood for the Good-Bad model with respect to the replica-averaged forward model data $f(\mathbf{X})$ using multiple data points.  Shown here, are three data points, two good data points $\{0.25, 0.3\}$ and one outlier $\{1.0\}$. The Good-Bad model ($\varphi=7.0$) is centered about the mean of the two good data points, demonstrating that this model can distinguish the good and bad data. The standard Gaussian likelihood ($\varphi=1$) is centered about the mean of all three data points. The colored curves are different values of nuisance parameter $\varphi$. Subplots (a) and (b) show how the Good-Bad model when $\varphi=1$ is equivalent to the Gaussian likelihood and harmonic potential energy function. Subplots (c) and (d) are the first and second derivatives of the potential energy curves shown in subplot (b).
}
  \label{fig:GB_derivatives}
\end{figure*}

\begin{figure*}
\centering
  \includegraphics[width=\linewidth]{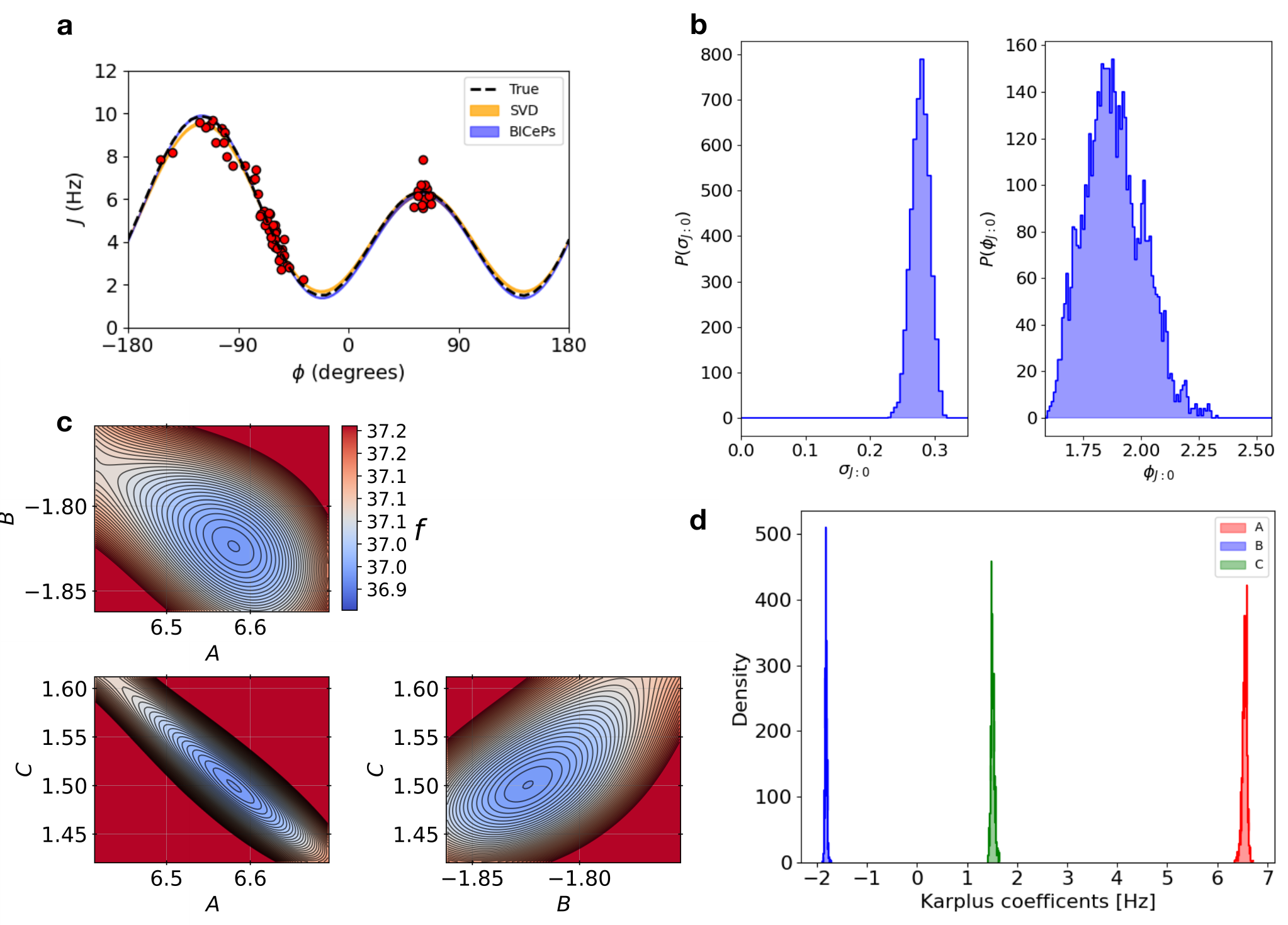}
  \caption{\small \textbf{BICePs predicted forward model parameters in the presence of random and systematic error ($\sigma_{data} = 0.471$ Hz) for a toy model system.}
  (a) Karplus curves predicted for SVD (orange) and BICePs (blue), where the "true" (black dashed line) parameters were set to be $\{A=6.51, B=-1.76, C=1.6 \}$. The extracted parameters from the SVD fitting were found to be $\{A=6.11 \pm 0.06, B=-1.63 \pm 0.04, C=1.80 \pm 0.04 \}$ and the BICePs was $\{A=6.6 \pm 0.037, B=-1.8 \pm 0.016, C=1.5 \pm 0.027\}$, averaged over three independent chains.  The uncertainty is represented by the thickness of the curves.
  When comparing the predicted J-couplings to the True J-couplings, the SVD approach yield an RMSE of $0.15\pm 0.02$ Hz and BICePs gives an RMSE of $0.06\pm 0.007$ Hz.
  For the BICePs calculation, we used the Good-Bad likelihood model with 32 replicas and burned for 20k steps, followed by 50k steps of additional sampling. Red dots correspond to the syntetic experimental J-coupling data points. (b) The marginal posterior distribution of uncertainty $p(\sigma_{J})$. The maximum a posteriori was determined to be $\hat{\sigma}_{J} = 0.272$ Hz, and the \textit{a posteriori} variance scaling parameter $\hat{\varphi}_{J} = 1.98$. (c) Landscapes of the BICePs score, $f$ for pairs of Karplus coefficients. (d) The marginal posterior distribution of Karplus coefficients.
}
  \label{fig:toy_example}
\end{figure*}

\begin{figure*}
\centering
  \includegraphics[width=0.6\linewidth]{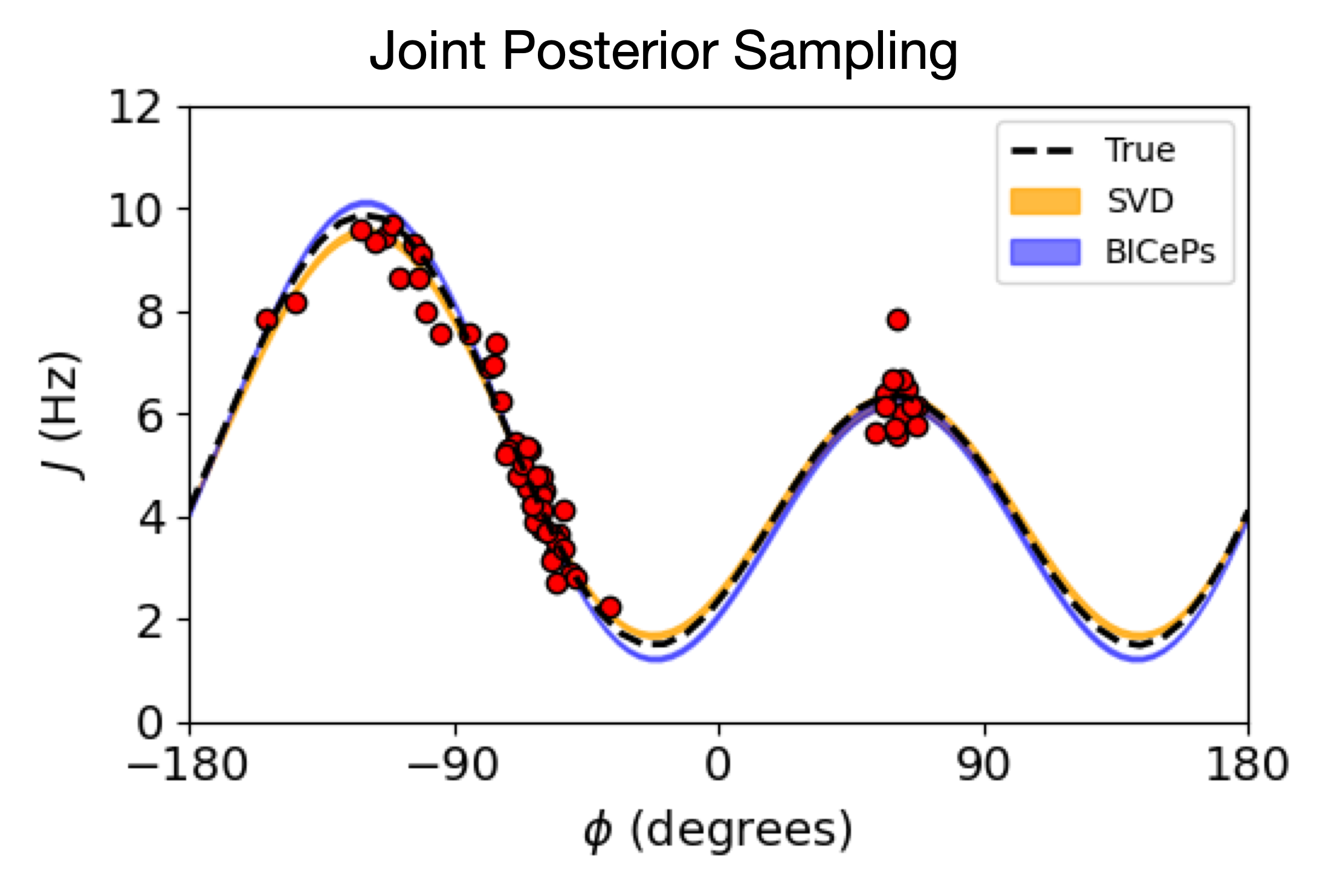}
  \caption{\small \textbf{The Student's model gives similar performance to the Good-Bad model.}  Karplus coefficients predicted using the Student's likeliood model ($\{A=6.8 \pm 0.033, B=-1.9 \pm 0.025, C=1.4 \pm 0.033 \}$) are compared against SVD when faced with random and systematic error ($\sigma_{data} = 0.471$ Hz). The "True" parameters were set to be $\{A=6.51, B=-1.76, C=1.6 \}$, the same as Figure \ref{fig:toy_example}.
  When comparing the predicted J-couplings to the True J-couplings, the SVD approach yield an RMSE of $0.15\pm 0.02$ Hz and BICePs gives an RMSE of $0.14\pm 0.011$ Hz.
}
  \label{fig:students_toy_example}
\end{figure*}

\begin{figure*}
\centering
  \includegraphics[width=0.6\linewidth]{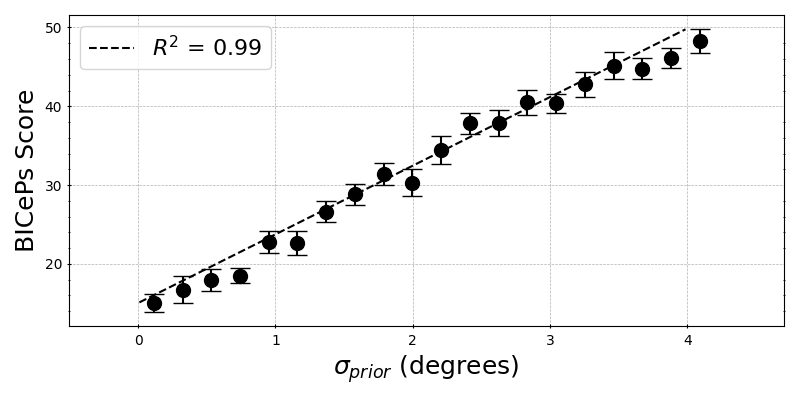}
  \caption{\small \textbf{The BICePs score is a measure of structural ensemble quality.} Using the same toy model system described in the main text. We vary the quality of the prior structural ensemble ($\sigma_{prior}$) by perturbing the "true" $\phi$ angles of the structural ensemble.  In these experiments, we induced over 1,000 random perturbations to the prior structural ensemble, and calculated BICePs scores for each. Error bars represent the standard deviation from the mean. The top panel shows the relationship between the BICePs score and the amount of error added to the structural ensemble. Each data point is an average across of 100 BICePs calculations. In these calculations, we used the Good-Bad likelihood model with 32 replicas.
}
  \label{fig:toy_biceps_score_corr}
\end{figure*}

\begin{figure*}
\begin{center}
\includegraphics[width=\linewidth]{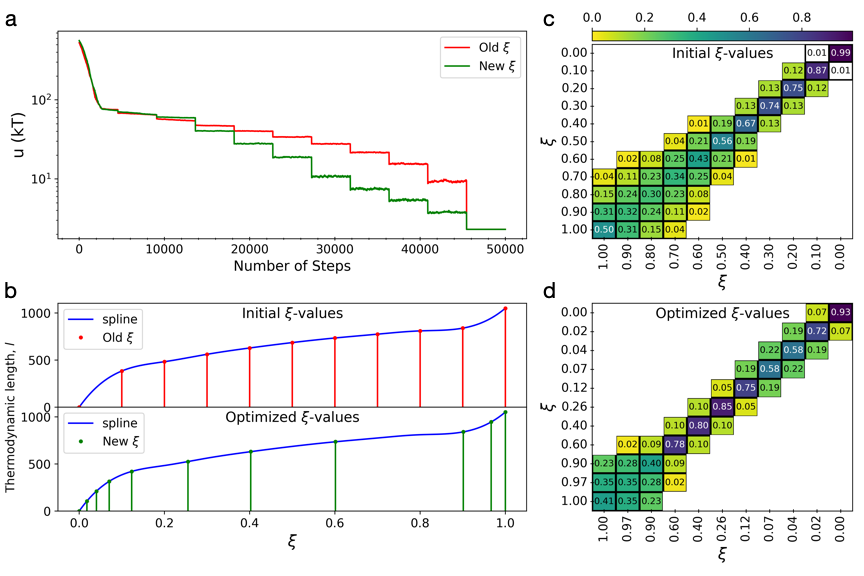}
\caption{\small \textbf{Optimization of thermodynamic intermediates $\xi_k$ used to calculate the BICePs score, $f_{\xi=0\rightarrow 1}$. }  In this scenario, sampling of the BICePs energy function $u = -\log p(\mathbf{X}, \sigma | D, \theta)$ is performed for a series of intermediates $\xi_k$ progressively from $\xi=1$ to $\xi=0$ (a).   The samples are used to optimize a new set of $\xi_k$ spaced uniformly in thermodynamic length (b).   While the thermodynamic overlap matrix for the unoptimized $\xi_k$ show poor overlap in the region of low $\xi$ values (c), the overlap matrix for the optimized $\xi_k$ shows good overlap for all neighboring intermediates (d). Reproduced fromRef.~\cite{novack2025simple} with permission.}
\label{fig:fmo_xi_optimization}
\end{center}
\end{figure*}

\begin{figure*}
\centering
  \includegraphics[width=0.6\linewidth]{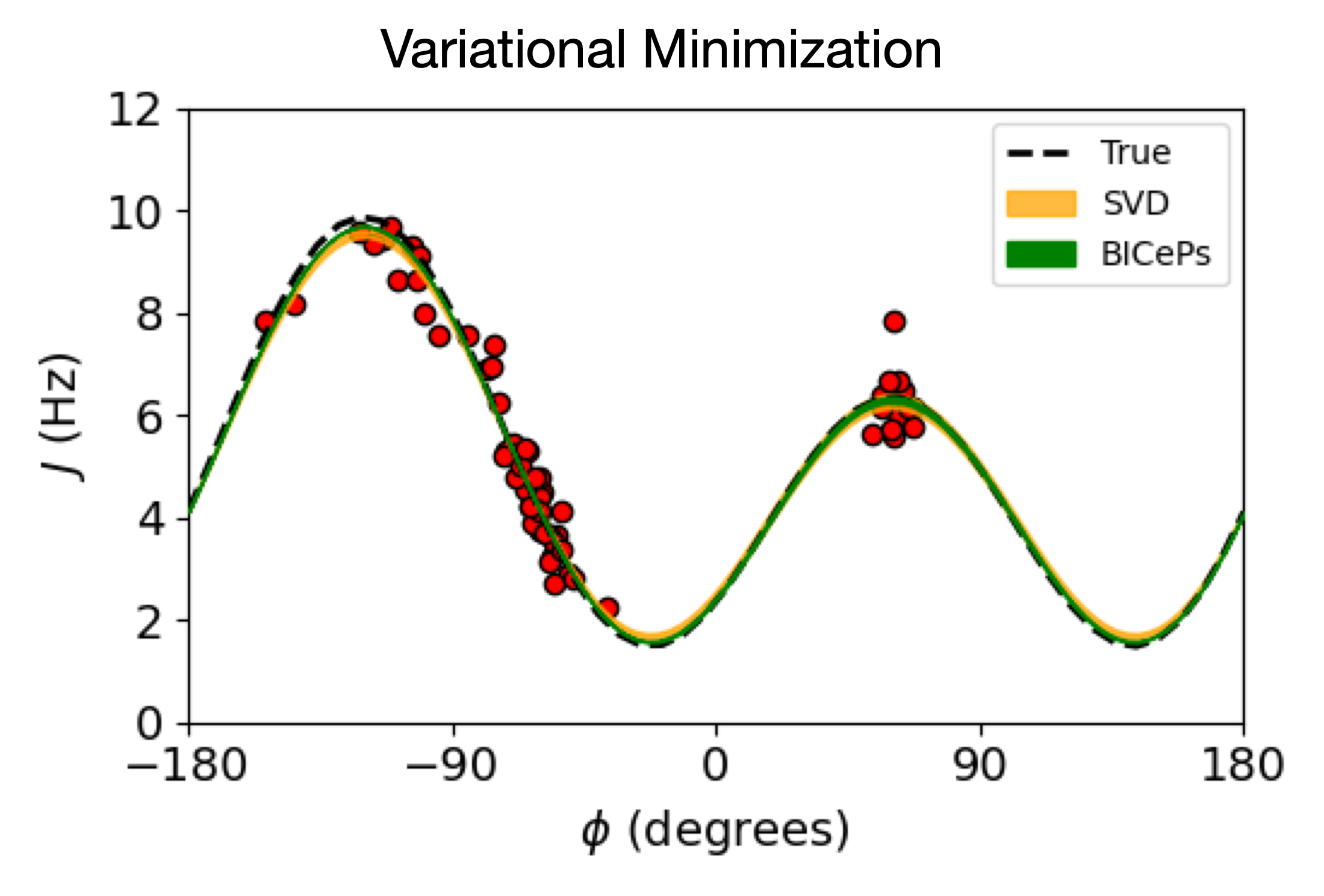}
  \caption{\small \textbf{Variational minimization of the BICePs score can be used to optimize forward model parameters.} By variational minimzation of the BICePs score, $f$ optimization traces converge to the "true" parameters. In these calculations, $\xi$-values were optimized prior to running the parameter refinement. We used the Good-Bad model with 4 replicas to minimize computational cost. Karplus coefficients predicted using the Good-Bad likeliood model ($\{6.31 \pm 0.02, -1.69 \pm 0.03, 1.69 \pm 0.01 \}$) are compared against SVD when faced with random and systematic error ($\sigma_{data} = 0.471$ Hz). The "True" parameters were set to be $\{A=6.51, B=-1.76, C=1.6 \}$, the same as Figure \ref{fig:toy_example}.
}
  \label{fig:variational_BS_toy_model}
\end{figure*}

\begin{figure*}
\centering
  \includegraphics[width=\linewidth]{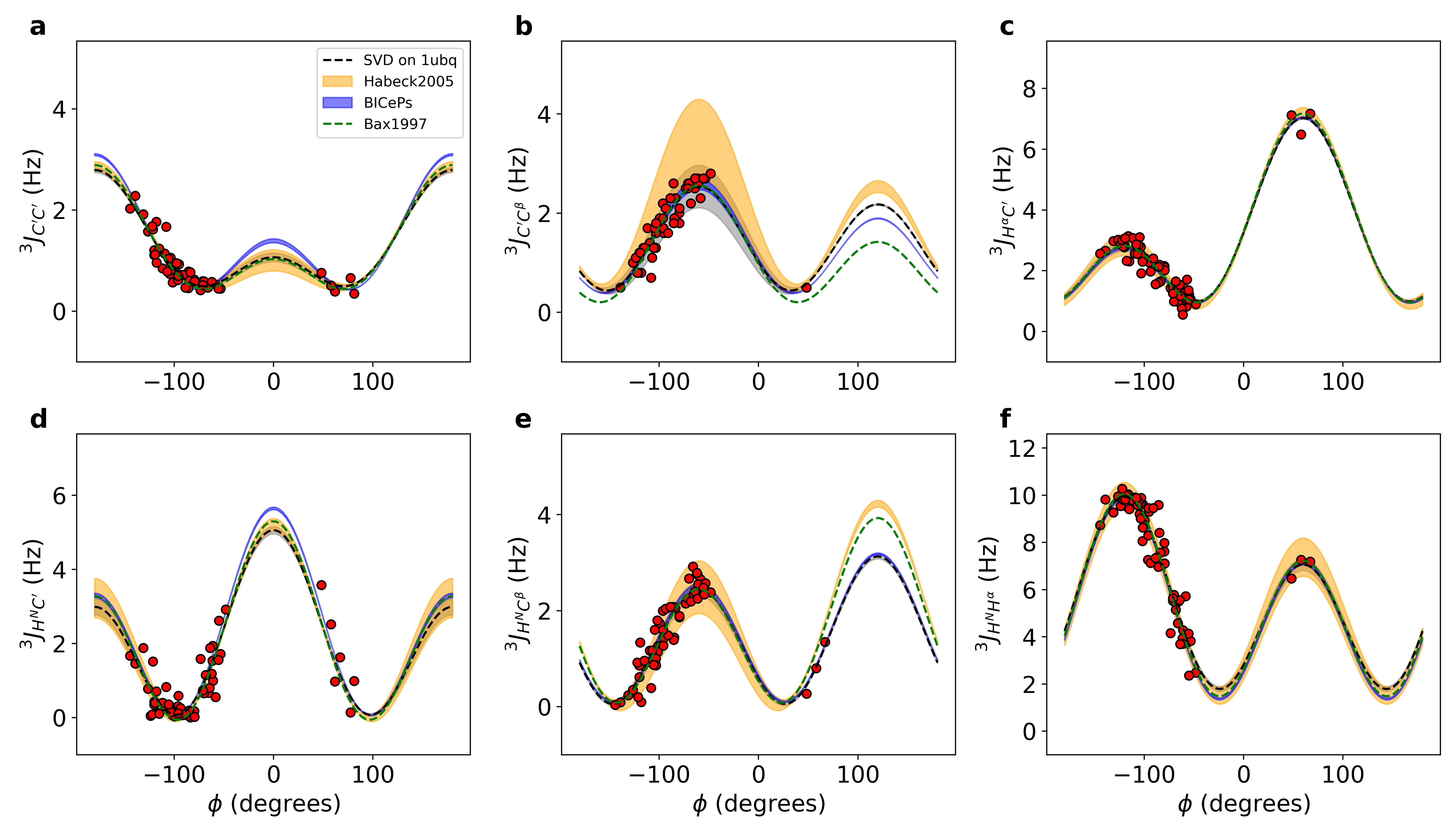}
  \caption{\small Karplus curves with BICePs refined Karplus coefficients using the 1d3z ensemble for (a-f) ${^{3}\!J}_{C^{\prime}C^{\prime}}$, ${^{3}\!J}_{C^{\prime}C^{\beta}}$, ${^{3}\!J}_{H^{\alpha} C^{\prime}}$,  ${^{3}\!J}_{H^{N} C^{\prime}}$,  ${^{3}\!J}_{H^{N} C^{\beta}}$, ${^{3}\!J}_{H^{N} H^{\alpha}}$. BICePs calculations were run using four chains with 32 replicas each, where we burned 50k steps, then sampled for another 50k MCMC steps.  For comparison, SVD on 1ubq using experimental scalar coupling constants with $\phi$-angles derived from the X-ray structure (black dashed line) and red dots correspond to the fitted data points. Additionally, parameterizations from Bax et al. 1997 (green), and parameterization from Habeck et al. 2005 (yellow) were overlaid for comparison. The thickness of the line corresponds to the uncertainty.}
  \label{fig:all_karplus_curves}
\end{figure*}

\begin{figure*}
\centering
  \includegraphics[width=\linewidth]{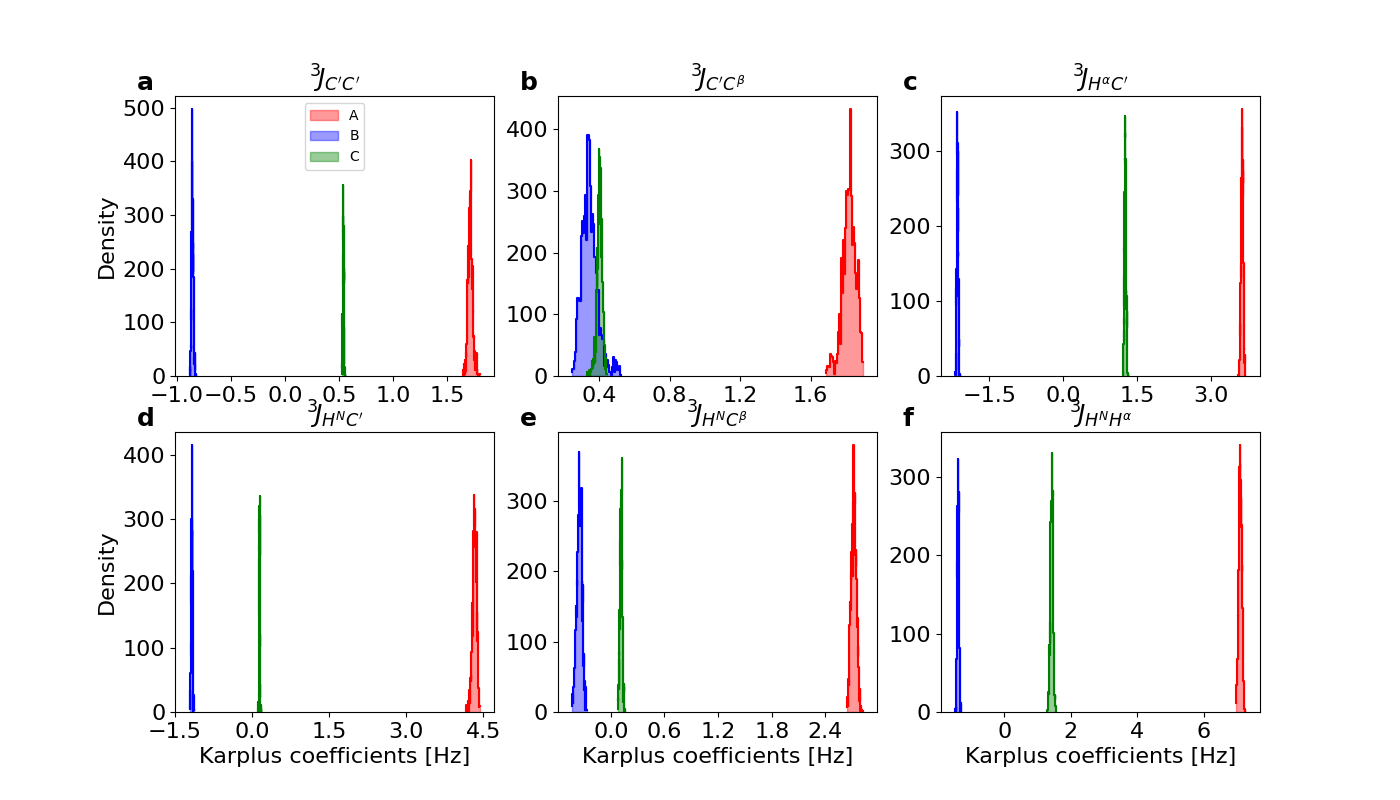}
  \caption{\small \textbf{Sampling the joint posterior distributions of six sets of Karplus coefficients using the Good-Bad model on the 1d3z ensemble.} BICePs calculations were run using four chains with 32 replicas each, where we burned 50k steps, then sampled for another 50k MCMC steps. These marginal posterior distributions shown here are from a randomly selected chain.
  }
  \label{fig:all_coefficient_histograms}
\end{figure*}

\begin{figure*}
\centering
  \includegraphics[width=\linewidth]{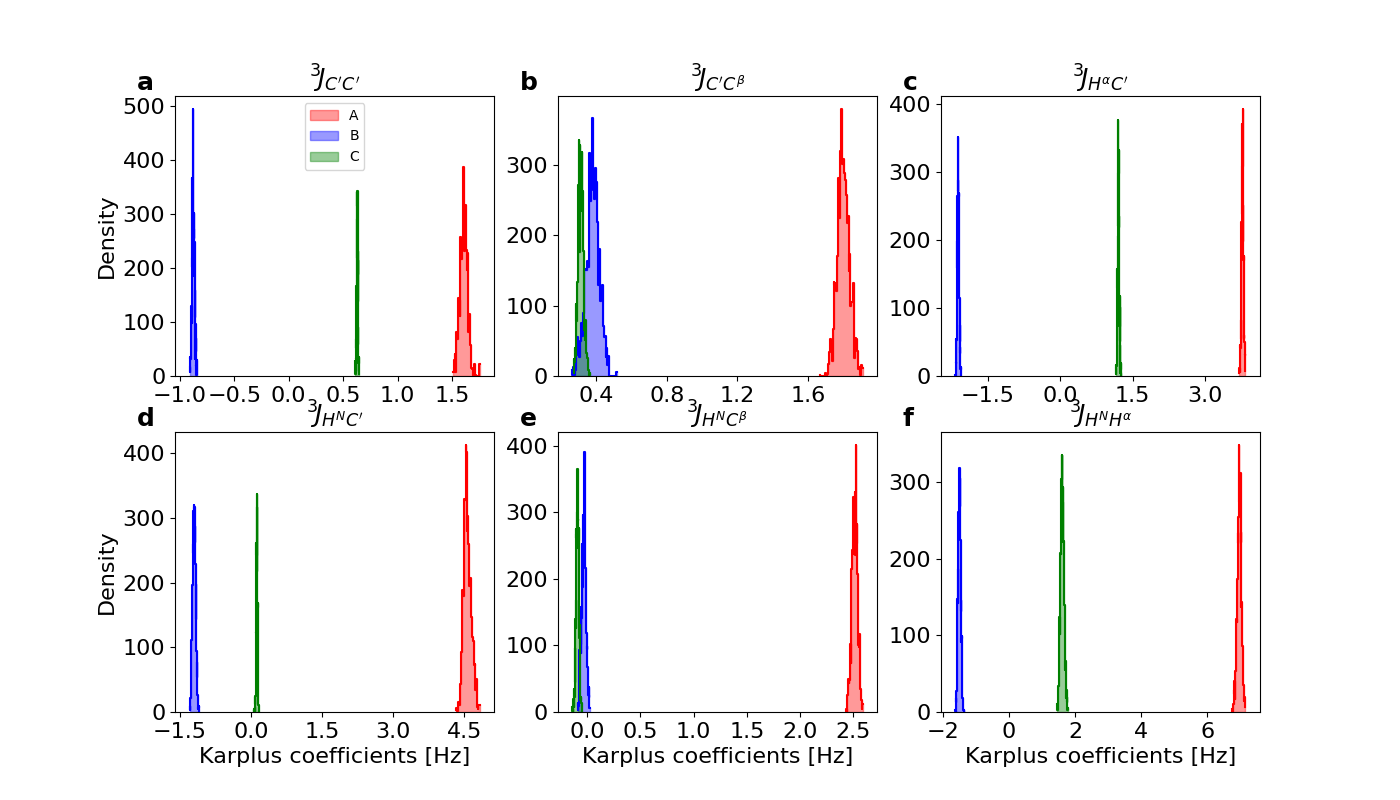}
  \caption{\small \textbf{Sampling the joint posterior distributions of six sets of Karplus coefficients using the Good-Bad model on the RosettaFold2 (RF2) ensemble.} BICePs calculations were run using four chains with 32 replicas each, where we burned 50k steps, then sampled for another 50k MCMC steps. Compare with Figure \ref{fig:all_coefficient_histograms} to see similarities.
  }
  \label{fig:all_coefficient_histograms_RF2}
\end{figure*}

\begin{figure*}
\centering
  \includegraphics[width=\linewidth]{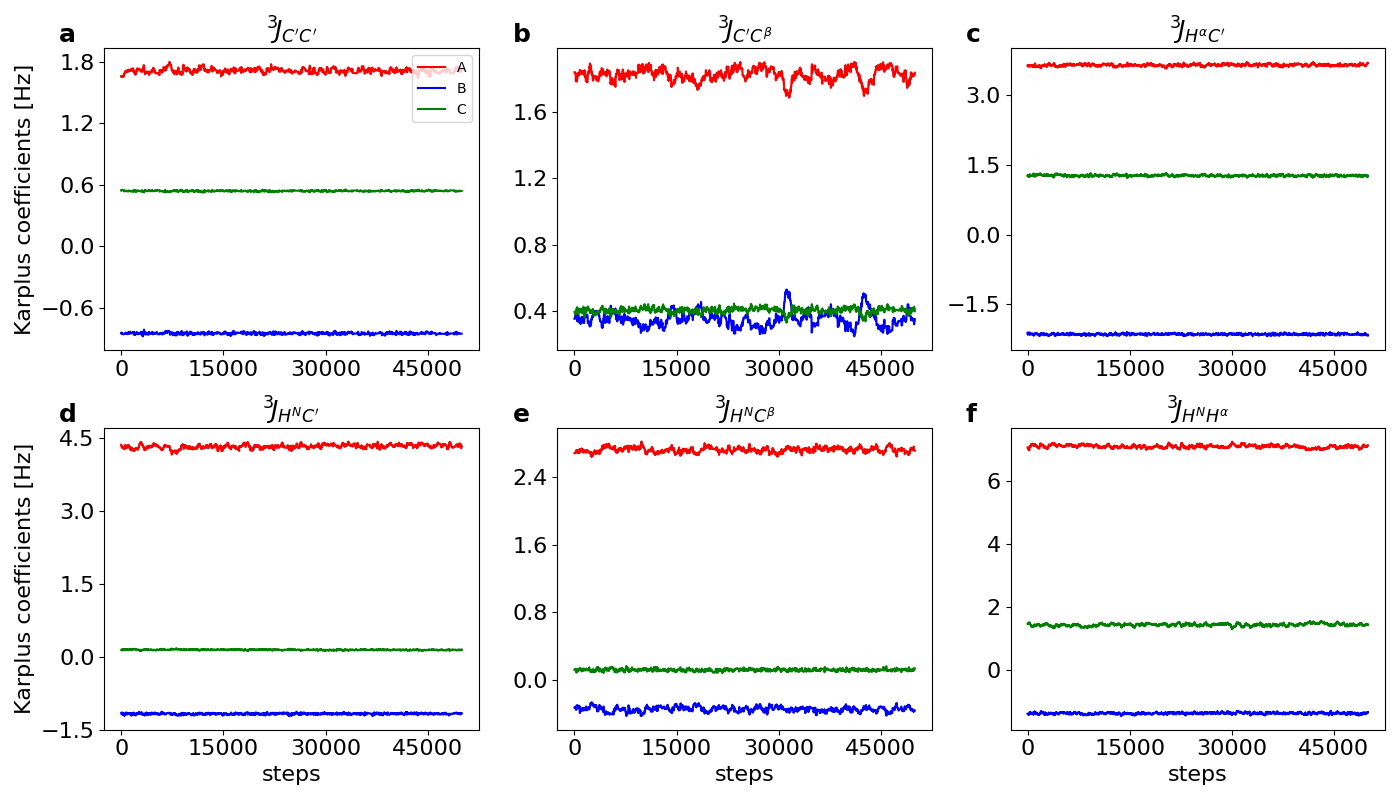}
  \caption{\small Traces of sampled Karplus coefficients for the 1d3z ensemble over 50k steps of MCMC, post-burn. BICePs calculations used the Good-Bad model with 32 replicas. Traces display low variance with no jumps, which demonstrates converged samples }
  \label{fig:all_coefficient_traces}
\end{figure*}

\begin{figure*}
\centering
  \includegraphics[width=0.8\linewidth]{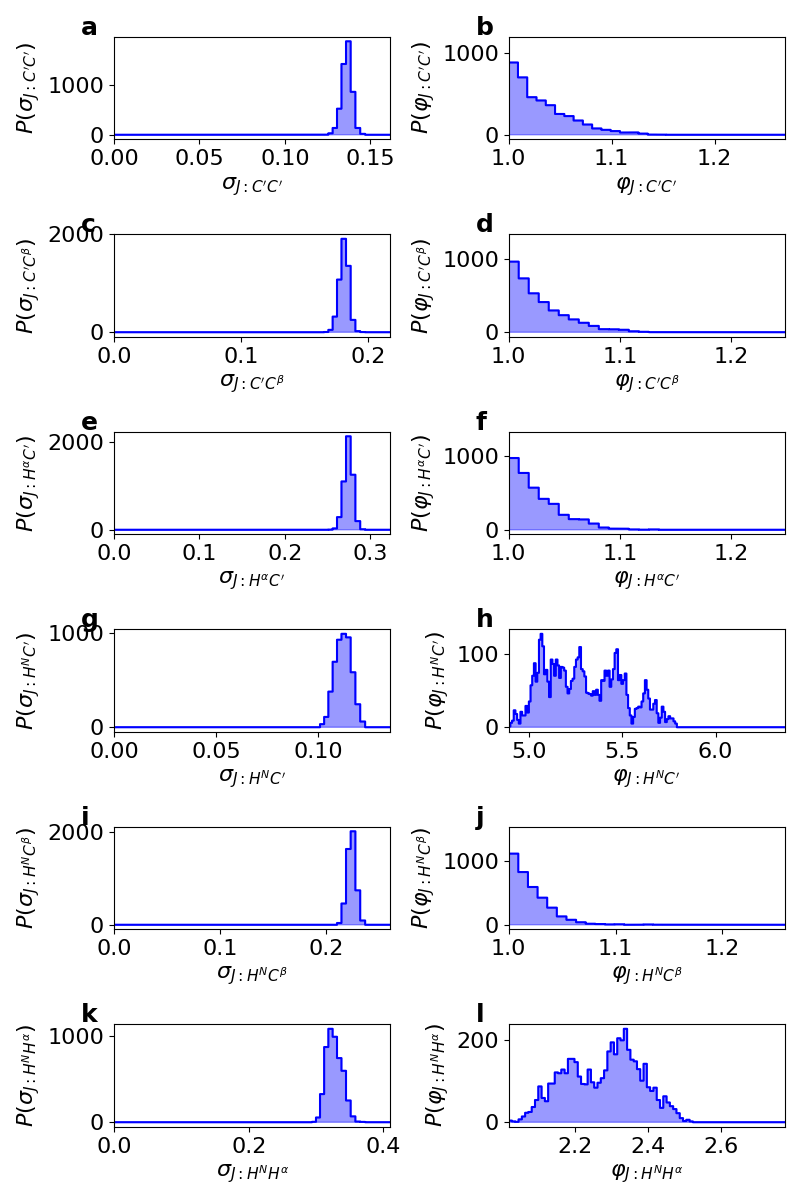}
  \caption{\small The marginal posterior distribution of $\sigma_{\text{J}}$, the uncertainty parameters for each set of J-coupling in the 1d3z ensemble. Densities are a result of posterior sampling of FM parameters during ensemble refinement using the Good-Bad model with 32 replicas.  The marginal posterior distributions of the variance scaling parameter $\varphi$ has a sampled mean slightly larger than 1.0 for particular sets of J-coupling, indicating that the functional form of the likelihood opted for long tails to account for a few outlier data points deviating from the mean. }
  \label{fig:marginal_distributions_of_sigma}
\end{figure*}

\begin{figure*}
\centering
   \includegraphics[width=0.95\linewidth]{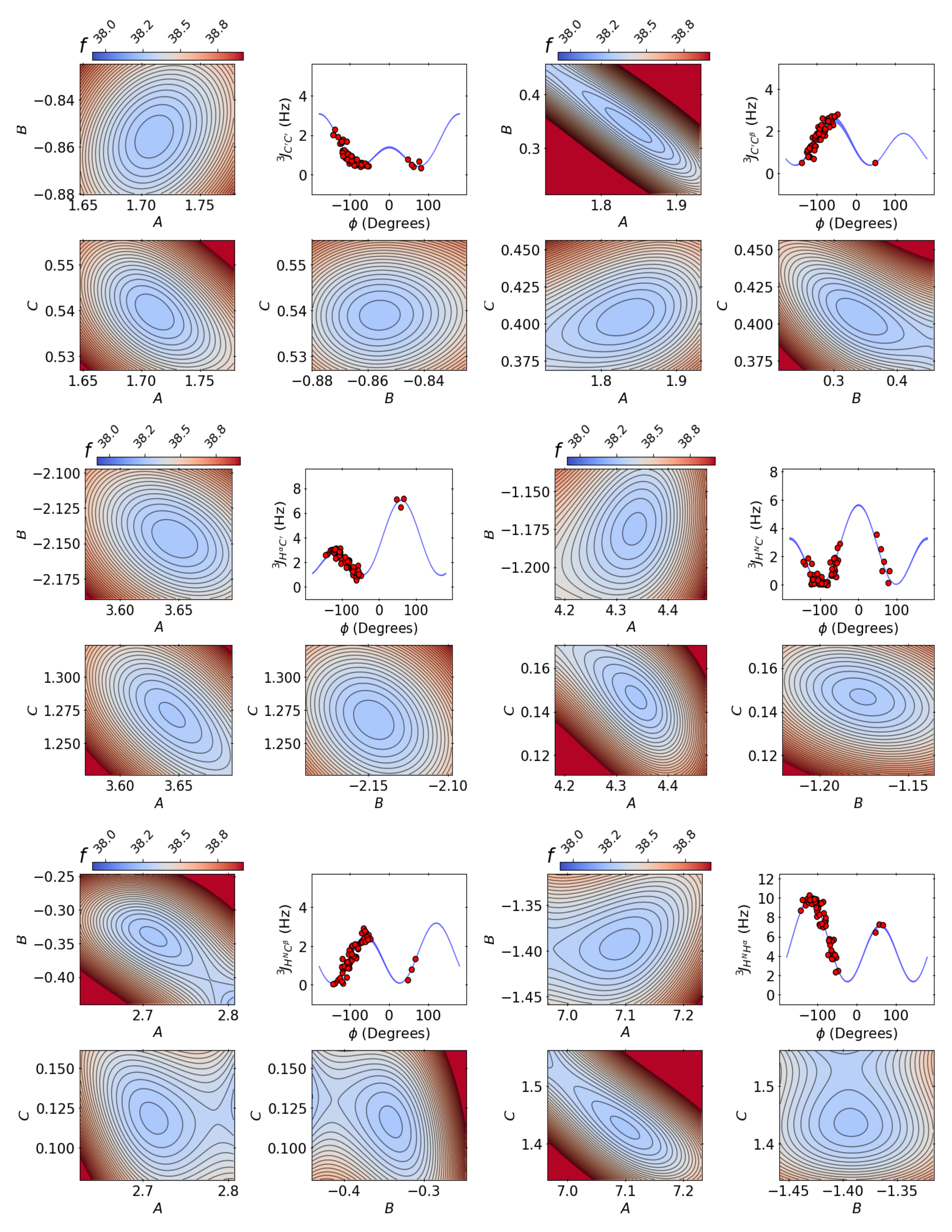}
  \caption{\small BICePs score landscapes of FM parameters on the 1d3z ensemble, unveiled during ensemble refinement. BICePs calcualtions used the Good-Bad model with 32 replicas. Each set of $\{A, B, C\}$ was included in the joint posterior of FM parameters.}
  \label{fig:all_landscapes}
\end{figure*}

%
%

\begin{figure*}
\centering
  \includegraphics[width=\linewidth]{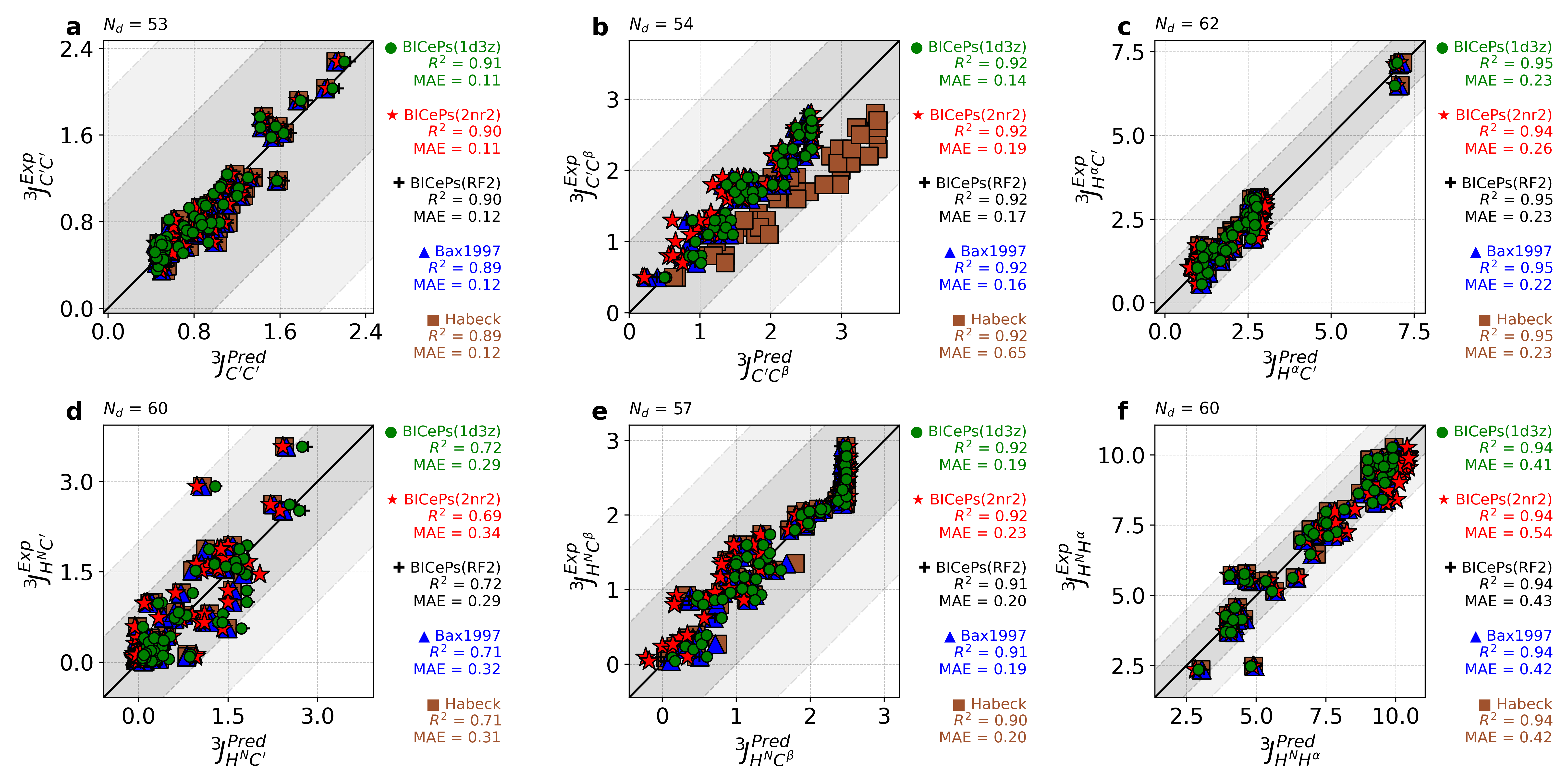}
  \caption{\small Validation of refined Karplus coefficients using BICePs on the 1D3Z structural ensemble show similar results to Bax1997 and minor improvements over Habeck2005 for scalar coupling predictions. Here, we compare models for predicting six sets of scalar coupling constants. Each panel shows strong correlations and relatively low error. }
  \label{fig:correlations_1d3z}
\end{figure*}

\begin{figure*}
\centering
  \includegraphics[width=\linewidth]{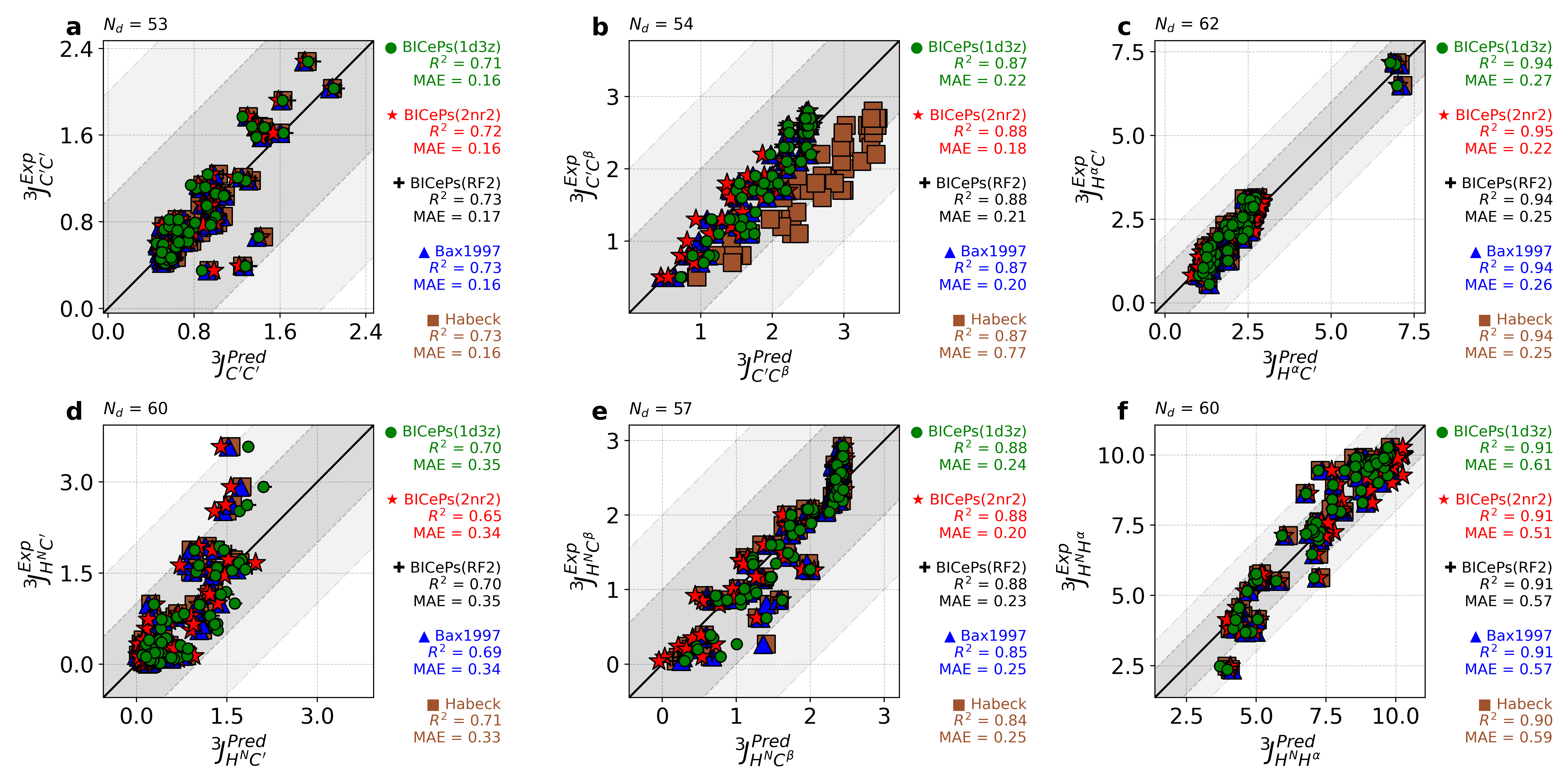}
  \caption{\small Validation of refined Karplus coefficients using BICePs on the 2NR2 structural ensemble show similar results to Bax1997 and minor improvements over Habeck2005 for scalar coupling predictions. Here, we compare various models for predicting six sets of scalar coupling constants. Each panel shows strong correlations between predictions and experiment. }
  \label{fig:correlations_2nr2}
\end{figure*}

\begin{figure*}
\centering
  \includegraphics[width=\linewidth]{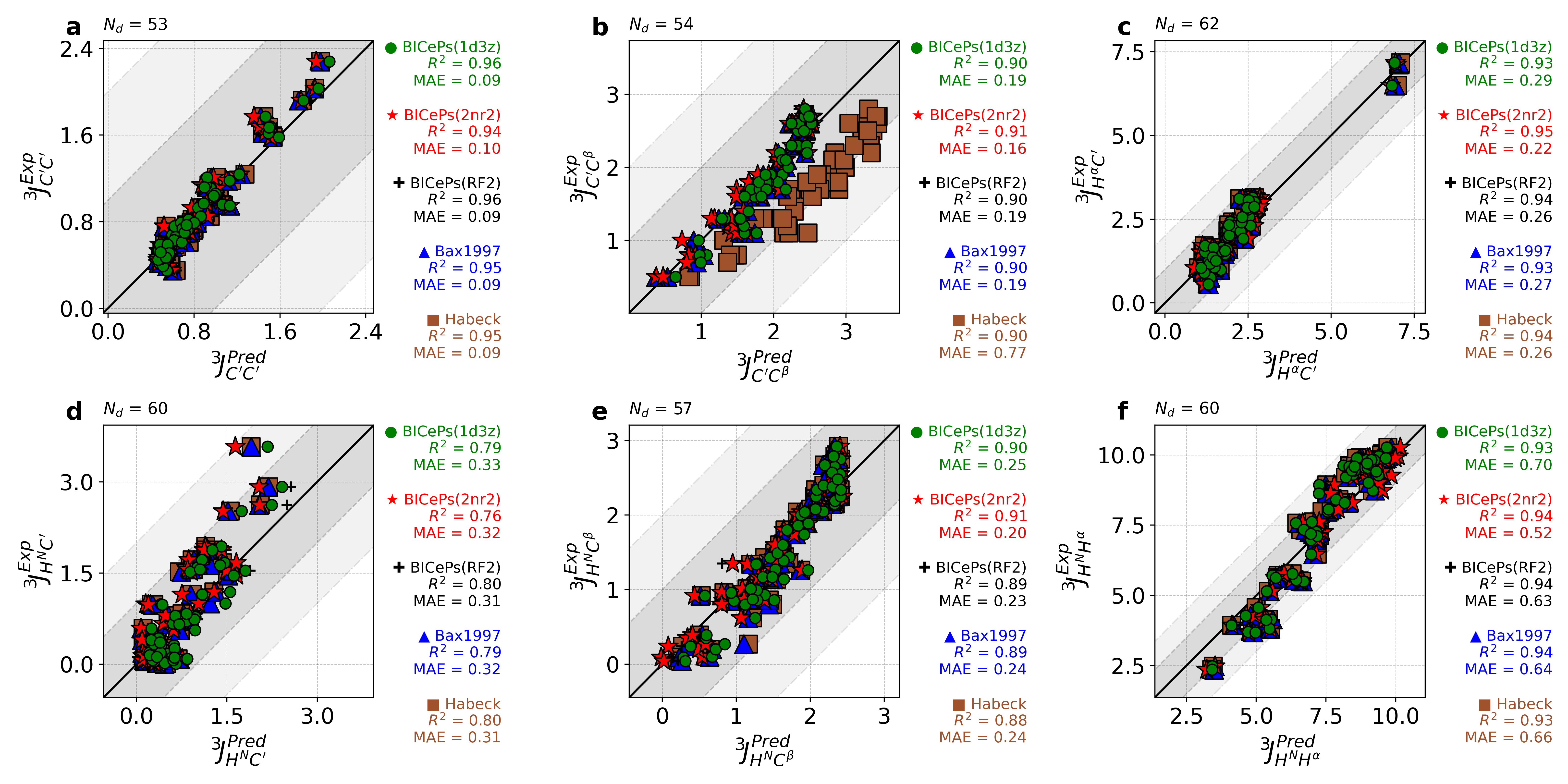}
  \caption{\small Validation of refined Karplus coefficients using BICePs on  the CHARMM22* structural ensemble show similar results to Bax1997 and minor improvements over Habeck2005 for scalar coupling predictions. Here, we compare various sets parameters for predicting six sets of scalar coupling constants. Each panel shows strong correlations between predictions and experiment. On average, BICePs parameters derived from the 2NR2 ensemble give the lowest MAE between experiment and predictions, whereas Habeck2005 has the highest due to ${^{3}\!J}_{C^{\prime}C^{\beta}}$.}
  \label{fig:correlations_charmm22*}
\end{figure*}

\begin{figure*}
\centering
  \includegraphics[width=\linewidth]{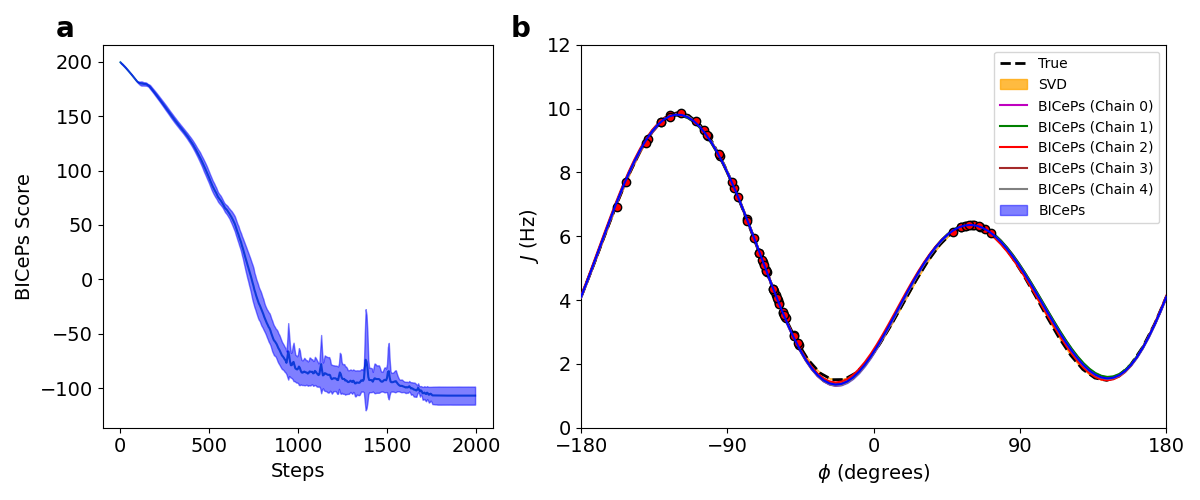}
  \caption{\small \textbf{BICePs trains NN parameters for $J$-coupling predictions using a toy model system.}
  (a) Minimization of the BICePs score during training of five independent neural networks (NNs). The solid blue line indicates the average BICePs score across the chains, with the shaded region representing the standard deviation. Convergence is evident by the plateau in score near 1500 training steps.
  (b) Karplus curves predicted by each of the five independently trained NNs. The mean prediction across all networks is shown in blue with uncertainty indicated by curve thickness (standard deviation across networks). The true curve (black dashed) corresponds to the parameters ${A = 6.51, B = -1.76, C = 1.60}$. Predictions obtained by singular value decomposition (SVD) are shown in orange, with extracted parameters ${A = 6.47 \pm 0.002, B = -1.75 \pm 0.001, C = 1.62 \pm 0.001}$. BICePs calculations were performed using a Good-Bad likelihood model and 32 replicas. No random or systematic noise was added to the synthetic data. Red dots correspond to the syntetic experimental J-coupling data points.
  When comparing the predicted J-couplings to the True J-couplings, the SVD approach yield an RMSE of $0.01\pm 0.001$ Hz and BICePs trained NN predictions have an RMSE of $0.04\pm 0.01$ Hz.
}
  \label{fig:toy_NN_example}
\end{figure*}

\begin{figure*}
\centering
  \includegraphics[width=\linewidth]{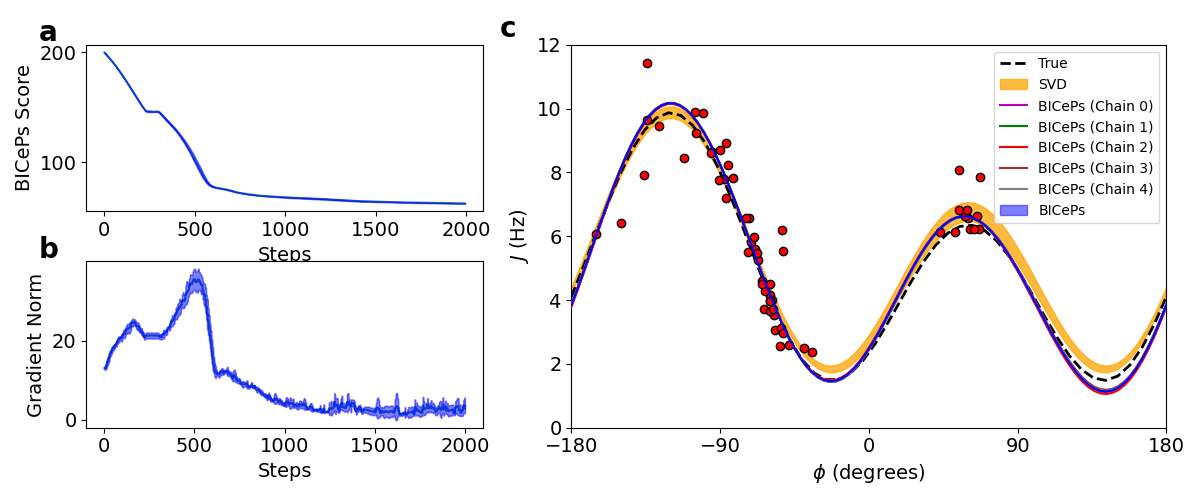}
  \caption{\small \textbf{BICePs trains NN parameters for $J$-coupling predictions using a toy model system in the presence of random and systematic error ($\sigma_{\text{data}} = 0.79$ Hz).}
  (a) Minimization of the BICePs score during training of five independent neural networks (NNs). The solid blue line indicates the average BICePs score across the chains, with the shaded region representing the standard deviation. Convergence is evident by the plateau in score near 1000 training steps.
  (b) The average gradient norm across the five chains during training.
  (c) Karplus curves predicted by each of the five independently trained NNs. The mean prediction across all networks is shown in blue with uncertainty indicated by curve thickness (standard deviation across networks). The true curve (black dashed) corresponds to the parameters ${A = 6.51, B = -1.76, C = 1.60}$. Predictions obtained by singular value decomposition (SVD) are shown in orange, with extracted parameters ${A = 6.39 \pm 0.13, B = -1.55 \pm 0.06, C = 1.94 \pm 0.10}$. BICePs calculations were performed using a Good-Bad likelihood model and 32 replicas. Red dots correspond to the syntetic experimental $J$-coupling data points.
  When comparing the predicted J-couplings to the True J-couplings, the SVD approach yield an RMSE of $0.24\pm 0.03$ Hz and BICePs trained NN predictions have an RMSE of $0.21\pm 0.01$ Hz.
}
  \label{fig:toy_NN_example_rand_sys_error}
\end{figure*}

\end{document}